\documentclass{aa}
\usepackage{color}
\usepackage{graphicx}
\usepackage{lscape}
\usepackage{natbib}
\usepackage[scaled]{helvet}
\usepackage[varg]{txfonts}
\usepackage{url}
\usepackage{xspace}
\bibpunct{(}{)}{;}{a}{}{,}
\newcounter{Rco}
\newcommand{\ionw}[3]{\mbox{\ion{#1}{#2}~$\lambda\,#3\,\mathrm{\AA}$}\xspace}
\newcommand{\ionww}[3]{\mbox{\ion{#1}{#2}~$\lambda\lambda\,#3\,\mathrm{\AA}$}\xspace}

\newcommand{\logg}{\mbox{$\log g$}\xspace}
\newcommand{\loggw}[1]{\mbox{$\log g\hspace{-0.5mm} =\hspace{-0.5mm}  #1$}}

\newcommand{\Teff}{\mbox{$T_\mathrm{eff}$}\xspace}
\newcommand{\Teffw}[1]{\mbox{$\Teff\hspace{-0.5mm} =\hspace{-0.5mm} #1 \,\mathrm{K}$}}
\newcommand{\ebv}{\mbox{$E_\mathrm{B-V}$}}

\newcommand{\Msol}{$M_\odot$\xspace}

\newcommand{\mmspr}{\hbox{}\hspace{+0.7cm}}

\newcommand{\gb}{\object{G191$-$B2B}\xspace}
\newcommand{\re}{\object{RE\,0503$-$289}\xspace}
\newcommand{\reb}{\object{RE\,0457$-$281}\xspace}
\newcommand{\wda}{\object{WD\,0458$-$303}\xspace}
\begin{document}

\title{Stellar laboratories }
\subtitle{VII. New \ion{Kr}{iv -- vii} oscillator strengths and an improved spectral analysis
           of the hot, hydrogen-deficient DO-type white dwarf \re
           \thanks
           {Based on observations with the NASA/ESA {\it Hubble} Space Telescope, obtained at the Space Telescope Science 
            Institute, which is operated by the Association of Universities for Research in Astronomy, Inc., under 
            NASA contract NAS5-26666.
           }$^,$
           \thanks
           {Based on observations made with the NASA-CNES-CSA Far Ultraviolet Spectroscopic Explorer.
           }$^,$
           \thanks
           {Based on observations made with ESO Telescopes at the La Silla Paranal Observatory under programme IDs 165.H-0588 and 167.D-0407. }$^,$
           \thanks
           {Based on observations obtained at the German-Spanish Astronomical Center,
           Calar Alto, operated by the Max-Planck-Institut f\"ur Astronomie Heidelberg
           jointly with the Spanish National Commission for Astronomy.}$^,$
           \thanks
           {Tables \ref{tab:kriv:loggf} to \ref{tab:krvii:loggf} are only available via the
            German Astrophysical Virtual Observatory (GAVO) service TOSS (http://dc.g-vo.org/TOSS).
           }
         }

\titlerunning{Stellar laboratories: New \ion{Kr}{iv -- vii} oscillator strengths}

\author{T\@. Rauch\inst{1}
        \and
        P\@. Quinet\inst{2,3}
        \and
        D\@. Hoyer\inst{1}
        \and
        K\@. Werner\inst{1}
        \and
        P\@. Richter\inst{4,5}
        \and
        J\@. W\@. Kruk\inst{6}
        \and
        M\@. Demleitner\inst{7}
        }

\institute{Institute for Astronomy and Astrophysics,
           Kepler Center for Astro and Particle Physics,
           Eberhard Karls University,
           Sand 1,
           72076 T\"ubingen,
           Germany \\
           \email{rauch@astro.uni-tuebingen.de}
           \and
           Physique Atomique et Astrophysique, Universit\'e de Mons -- UMONS, 7000 Mons, Belgium
           \and
           IPNAS, Universit\'e de Li\`ege, Sart Tilman, 4000 Li\`ege, Belgium
           \and
           Institute of Physics and Astronomy, University of Potsdam, Karl-Liebknecht-Str. 24/25, 14476 Golm, Germany
           \and
           Leibniz-Institut für Astrophysik Potsdam (AIP), An der Sternwarte 16, 14482 Potsdam, Germany            
           \and
           NASA Goddard Space Flight Center, Greenbelt, MD\,20771, USA
           \and
           Astronomisches Rechen-Institut (ARI), Centre for Astronomy of Heidelberg University, M\"onchhofstra\ss e 12-14, 69120 Heidelberg, Germany}

\date{Received 14 January 2016; accepted 22 February 2016}

\abstract {For the spectral analysis of high-resolution and high signal-to-noise (S/N) spectra of hot stars,
           state-of-the-art non-local thermodynamic equilibrium (NLTE) 
           model atmospheres are mandatory. These are strongly
           dependent on the reliability of the atomic data that is used for their calculation.
          }
          {New \ion{Kr}{iv-vii} oscillator strengths for a large number of lines enable us to construct
           more detailed model atoms for our NLTE
           model-atmosphere calculations. This enables us to search for additional Kr lines in
           observed spectra and to improve Kr abundance determinations.
          }
          {We calculated \ion{Kr}{iv-vii} oscillator strengths
           to consider radiative and collisional bound-bound transitions
           in detail in our NLTE stellar-atmosphere models
           for the analysis of Kr lines that are  exhibited in
           high-resolution and high S/N ultraviolet (UV) observations of 
           the hot white dwarf \re.
          }
          {We reanalyzed the effective temperature and surface gravity and determined
           \Teffw{70\,000 \pm 2000} and $\log\,(g\,/\,\mathrm{cm\,s^{-2}}) = 7.5 \pm 0.1$.
           We newly identified ten \ion{Kr}{v} lines and one \ion{Kr}{vi} line in the spectrum of \re.
           We measured a Kr abundance of $-3.3 \pm 0.3$ (logarithmic mass fraction).
           We discovered that the interstellar absorption toward \re has a multi-velocity
           structure within a radial-velocity interval of 
           $-40\,\mathrm{km\,s^{-1}} < v_\mathrm{rad} < +18\,\mathrm{km\,s^{-1}}$.
           }
          {Reliable measurements and calculations of atomic data are a prerequisite for
           state-of-the-art NLTE stellar-atmosphere modeling. 
           Observed \ion{Kr}{v-vii} line profiles in the UV spectrum of the white dwarf \re 
           were simultaneously well reproduced with our newly calculated oscillator strengths. 
          }

\keywords{atomic data --
          line: identification --
          stars: abundances --
          stars: individual: \re\ --
          stars: individual: \reb\ --
          virtual observatory tools
         }

\maketitle

\section{Introduction}
\label{sect:intro}

Reliably determining the abundance  of trans-iron elements in hot white dwarf (WD) stars,
e.g., 
\gb and \re
\citep[\object{WD\,0501+527} and \object{WD\,0501$-$289}, respectively),][]{mccooksion1999,mccooksion1999cat},
recently became possible with the calculation of transition probabilities for highly 
ionized 
Zn (atomic number $Z = 30$), 
Ga (31), 
Ge (32), 
Mo (42), and 
Ba (56) 
\citep{rauchetal2014zn, rauchetal2015ga, rauchetal2012ge, rauchetal2016mo, rauchetal2014ba}. 
These analyses were initiated by the discovery of lines
of 
Ga, 
Ge, 
As (33), 
Se (34), 
Kr (36), 
Mo, 
Sn (50), 
Te (52), 
I, (53), and 
Xe (54) in the Far Ultraviolet Spectroscopic Explorer (FUSE) 
spectrum of the hydrogen-deficient DO-type WD \re by \citet{werneretal2012}. Owing to the lack
of atomic data at that time, they could only measure  the Kr and Xe abundances 
($-4.3 \pm 0.5$ and $-4.2 \pm 0.6$ in logarithmic mass fractions, respectively).
We calculated new \ion{Kr}{iv-vii} transition probabilities to construct more
detailed model atoms that are used in our non-local thermodynamical equilibrium (NLTE) model-atmosphere
calculations to improve the Kr abundance determination.

In Sects.\,\ref{sect:observation} and \ref{sect:models}, we briefly describe the available observations,
our model-atmosphere code, and the atomic data. Details of the transition-probability calculations and a
comparison of the results with literature values are given in Sect.\,\ref{sect:krtrans}. 
Based on our state-of-the-art NLTE models, we start our spectral analysis with a verification of
the previous determination of the effective temperature and surface gravity by
\citet[][\Teffw{70\,000}, $\log\,(g\,/\,\mathrm{cm\,s^{-2}}) = 7.5$]{dreizlerwerner1996} in Sect.\,\ref{sect:tefflogg}.
An improved Kr abundance analysis for \re is then presented in Sect.\,\ref{sect:abundance}.
The stellar mass and the distance of \re are revisited in Sect.\,\ref{sect:massdistance}.
At the end (Sect.\,\ref{sect:ism}), we take a look at the velocity field of the observed interstellar line absorption
and compare it with that of the nearby hydrogen-rich, DA-type WD 
\reb \citep[\object{WD\,0455$-$282}, $m_\mathrm{V} = 13.90$][]{mccooksion1999,mccooksion1999cat,gianninasetal2011}.
We summarize and conclude in Sect.\,\ref{sect:results}.

\section{Observations}
\label{sect:observation}

We analyzed ultraviolet (UV) FUSE \citep[described in detail by][]{werneretal2012}
and HST/STIS ({\it Hubble} Space Telescope / Space Telescope Imaging Spectrograph)
observations  ($1144\,\mathrm{\AA} < \lambda <  3073\,\mathrm{\AA}$) of \re, that were performed on 2014 08 14. 
The latter spectrum was co-added from two observations with grating E140M
(exposure times 2493\,s and 3001\,s, $1144\,\mathrm{\AA} - 1709\,\mathrm{\AA}$, resolving power $R \approx 45\,800$), and
two observations with grating E230M 
(1338\,s, $1690\,\mathrm{\AA} - 2366\,\mathrm{\AA}$ and 
 1338\,s, $2277\,\mathrm{\AA} - 3073\,\mathrm{\AA}$, $R \approx 30\,000$).
These STIS observations are retrievable from the Barbara A\@. Mikulski Archive for Space Telescopes (MAST).

In addition to the UV observations, we used optical spectra that were obtained at the European Southern Observatory (ESO)
and the Calar Alto (CA) observatory.
In the framework of the Supernova Ia Progenitor Survey project \citep[SPY,][]{napiwotzkietal2001, napiwotzkietal2003},
observations were performed on 2000 09 09 and 2001 04 08 with the Ultraviolet and Visual 
Echelle Spectrograph (UVES) attached to the Very Large Telescope (VLT) located at ESO.
The co-added spectra cover the wavelength intervals [3290\,\AA, 4524\,\AA],  [4604\,\AA, 5609\,\AA], and [5673\,\AA, 6641\,\AA]
with a resolution of about 0.2\,\AA. 
Two spectra [4094\,\AA, 4994\,\AA] and  [5680\,\AA, 6776\,\AA] were taken with the Cassegrain TWIN Spectrograph that was attached to the 
3.5\,m telescope at the CA observatory. Their resolution is about 3\,\AA\ \citep[the same spectra were used by][]{dreizlerwerner1996}.

For \reb, we used FUSE spectra  
(ObsIds P1041101000, P1041102000, and P1041103000 from 2000 02 03, 2000 02 04, and 2000 02 07, respectively, with a total observation time 
of 47\,465\,s) that were obtained with the medium-resolution (MDRS) aperture. In addition, we used an IUE 
(International Ultraviolet Explorer) spectrum ([1153\,\AA, 1947\,\AA])
that was co-added from four observations that were obtained in high-resolution ($R \approx 10\,000$) mode
with the large aperture 
\citep[Data Ids SWP46302, SWP56213, SWP56262, and SWP56267 from 1992-11-19, 1995-11-18, 1995-12-02, and 1995-12-04, respectively, with a
total exposure time of 168\,360\,s,][]{holbergetal1998}.                
This is available via the MAST High-Level Science Products.

If not otherwise explicitly mentioned, all synthetic spectra shown in this paper, which are compared 
with an observation, are convolved with a Gaussian to model the respective resolving power. 
The observed spectra are shifted to rest wavelengths according to our measurement of 
the radial velocity $v_\mathrm{rad} = 25.5\,\mathrm{km\,s^{-1}}$.

\section{Model atmospheres and atomic data}
\label{sect:models}

We calculated plane-parallel, chemically homogeneous model-atmospheres in hydrostatic and radiative 
equilibrium with our T\"ubingen NLTE Model Atmosphere Package
\citep[TMAP\footnote{\url{http://astro.uni-tuebingen.de/~TMAP}},][]{werneretal2003,tmap2012}.
Model atoms were provided by the T\"ubingen Model Atom Database
\citep[TMAD\footnote{\url{http://astro.uni-tuebingen.de/~TMAD}},][]{rauchdeetjen2003}. TMAD
was constructed as part of the T\"ubingen contribution to the German Astrophysical Virtual Observatory 
(GAVO\footnote{\url{http://www.g-vo.org}}).

Our Kr model atom was designed with a statistical method \citep[similar to][]{rauchetal2015ga}
by calculating the so-called super levels and super lines with our Iron Opacity and Interface
\citep[IrOnIc\footnote{\url{http://astro.uni-tuebingen.de/~TIRO}},][]{rauchdeetjen2003}.
Using our approach, we neglected spin system and parity of the individual levels in the calculation
of the super levels. This is justified because, in the atmosphere of \re, the deviation of the levels' occupation numbers
from LTE is small in the line-forming region of the atmosphere. The detailed fits of our theoretical 
line profiles to the observations (Sect.\,\ref{sect:abundance}) do no give any hint of inconsistencies.
To process our new Kr data, we transferred it into Kurucz's format\footnote{\url{http://kurucz.harvard.edu/atoms.html}},
which is readable by IrOnIc. The statistics of our Kr model atom are summarized in Table\,\ref{tab:ironic}.

\begin{table}\centering
\caption{Statistics of \ion{Kr}{iv - vii} atomic levels and line transitions from
         Tables\,\ref{tab:kriv:loggf}-\ref{tab:krvii:loggf}, respectively.}         
\label{tab:ironic}
\begin{tabular}{ccccc}
\hline
\hline
Ion       & Atomic levels & Lines & Super levels & Super lines \\
\hline
\sc{iv}   &            83 &   911 &            7 &          19 \\
\sc{v}    &            64 &   553 &            7 &          16 \\
\sc{vi}   &            69 &   843 &            7 &          19 \\
\sc{vii}  &            70 &   743 &            7 &          21 \\
\hline
          &           286 &  3050 &           28 &          75 \\
\hline
\end{tabular}
\end{table}  

For the calculation of cross-sections, we followed \citet{rauchdeetjen2003} for the
transition types\\
-- collisional bound-bound: \citet{vanregemorter1962} formula for known f-values and an
approximate formula for unknown f-values,\\
-- radiative bound-bound: approximate formula by \citet{cowley1970,cowley1971} for the quadratic Stark effect, and\\
-- collisional and radiative bound-free: \citet{seaton1962} formula with hydrogen-like threshold cross-sections.

For Kr and all other species, level dissolution (pressure ionization) following
\citet{hummermihalas1988} and \citet{hubenyetal1994} is accounted for. 
Stark broadening tables of \citet{bcs1974} are used for \ionw{He}{i}{4471} and of
\citet{schoeningbutler1989} for \ion{He}{ii} lines.

\section{Oscillator-strength calculations in krypton ions}
\label{sect:krtrans}

New oscillator strengths were computed for transitions in \ion{Kr}{iv-vii} ions in the present work. 
The method used was the same as the one considered in our previous studies that focused on Zn, Ga, Ge, Mo, and Ba ions 
\citep{rauchetal2014zn, rauchetal2015ga, rauchetal2012ge, rauchetal2016mo, rauchetal2014ba},
namely the Relativistic Hartree-Fock (HFR) method \citep{cowan1981} that was modified 
to take core-polarization effects into account (HFR+CPOL), as described by \citet{quinetetal1999,quinetetal2002}. 
In each Kr ion, the same core-polarization parameters were used, i.e., a dipole polarizability 
$\alpha_{\mathrm d} = 0.20\,{\mathrm a_0^3}$ 
and a cut-off radius 
$r_{\mathrm c} = 0.55\,{\mathrm a_0}$. 
The former value, taken from \citet{johnsonetal1983}, corresponds to a Kr$^{8+}$ 
closed-shell ionic core of the type 
1s$^2$2s$^2$2p$^6$3s$^2$3p$^6$3d$^{10}$,
while the latter value was chosen as the mean value of 
$\left< r\right> $ for the outermost core orbital (3d), as calculated by the HFR approach. Intravalence correlations were 
considered by explicitly including the following configurations in the physical models:\\
\noindent{\sc\bf \ion{Kr}{iv}\,\,}
4s$^2$4p$^3$+ 
4s$^2$4p$^2$5p + 
4s$^2$4p$^2$6p + 
4s$^2$4p$^2$4f + 
4s$^2$4p$^2$5f + 
4s$^2$4p$^2$6f + 
4s4p$^3$4d + 
4s4p$^3$5d + 
4s4p$^3$6d + 
4s4p$^3$5s + 
4s4p$^3$6s + 
4s$^2$4p4d$^2$ + 
4s$^2$4p4f$^2$ + 
4p$^5$ + 
4p$^4$4f (odd parity) and 
4s4p$^4$ + 
4s$^2$4p$^2$4d + 
4s$^2$4p$^2$5d + 
4s$^2$4p$^2$6d + 
4s$^2$4p$^2$5s + 
4s$^2$4p$^2$6s + 
4s4p$^3$4f+ 
4s4p$^3$5f+ 
4s4p$^3$6f+ 
4s4p$^3$5p+ 
4s4p$^3$6p+ 
4p$^4$4d + 
4p$^4$5s (even parity), \\
\noindent{\sc\bf \ion{Kr}{v}\,\,}
4s$^2$4p$^2$ + 
4s$^2$4p5p + 
4s$^2$4p6p + 
4s$^2$4p4f + 
4s$^2$4p5f + 
4s$^2$4p6f + 
4s4p$^2$4d + 
4s4p$^2$5d + 
4s4p$^2$6d + 
4s4p$^2$5s + 
4s4p$^2$6s + 
4s$^2$4d$^2$ + 
4s$^2$4f$^2$ + 
4p$^4$ + 
4p$^3$4f (even parity) and 
4s4p$^3$ + 
4s$^2$4p4d + 
4s$^2$4p5d + 
4s$^2$4p6d + 
4s$^2$4p5s + 
4s$^2$4p6s + 
4s4p$^2$4f + 
4s4p$^2$5f + 
4s4p$^2$6f + 
4s4p$^2$5p + 
4s4p$^2$6p + 
4p$^3$4d + 
4p$^3$5s (odd parity), \\
\noindent{\sc\bf \ion{Kr}{vi}\,\,}
4s$^2$4p + 
4s$^2$5p + 
4s$^2$6p + 
4s$^2$4f + 
4s$^2$5f + 
4s$^2$6f + 
4s4p4d + 
4s4p5d + 
4s4p6d + 
4s4p5s + 
4s4p6s + 
4p$^3$ + 
4p$^2$4f + 
4s4d4f + 
4p4d$^2$ + 
4d$^2$4f + 
4p4f$^2$ (odd parity) and 
4s4p$^2$ + 
4s$^2$4d + 
4s$^2$5d + 
4s$^2$6d + 
4s$^2$5s + 
4s$^2$6s + 
4s4p5p + 
4s4p6p + 
4s4p4f + 
4s4p5f + 
4s4p6f + 
4p$^2$4d + 
4p$^2$5s + 
4s4d5s + 
4s4d$^2$ + 
4s4f$^2$ (even parity), \\
\noindent{\sc\bf \ion{Kr}{vii}\,\,}
4s$^2$ + 
4p$^2$ + 
4d$^2$ + 
4f$^2$ + 
5s$^2$ + 
4s4d + 
4s5d + 
4s6d + 
4s5s + 
4s6s + 
4p4f + 
4p5f + 
4d5s + 
4p5p (even parity) and 
4s4p + 
4s5p + 
4s6p + 
4s4f + 
4s5f + 
4s6f + 
4p5s + 
4p4d + 
4p5d + 
4d4f (odd parity).

The HFR+CPOL method was then combined with a semi-empirical optimization of the radial parameters 
to minimize the discrepancies between calculated and experimental energy levels. More precisely, the energy levels reported by \citet{saloman2007} were used in \ion{Kr}{iv}
to adjust the radial parameters corresponding to the 
4p$^3$, 
4p$^2$5p, 
4s4p$^4$, 
4p$^2$4d, 
4p$^2$5d, 
4p$^2$5s, and 
4p$^2$6s 
configurations. In \ion{Kr}{v}, the experimental level values taken from \citet{saloman2007} and \citet{rezendeetal2010}
were included in the adjustment of some parameters in the 
4p$^2$, 
4p5p, 
4s4p$^2$4d, 
4p$^4$, 
4s4p$^3$, 
4p4d, 
4p5d, 
4p5s, and 
4p6s 
configurations. In the case of \ion{Kr}{vi}, the energy levels from \citet{saloman2007} and \citet{fariasetal2011} enabled us to 
optimize the radial parameters by describing the 
4p,
5p, 
4s4p4d, 
4p$^3$, 
4s4p$^2$, 
4s$^2$4d, 
4s$^2$5s, 
4s4p4f, 
4s4p5p, and 
4p$^2$4d 
configurations. Finally, the energy levels measured by \citet{rainerietal2014} were used to fit the parameters of the 
4s$^2$, 
4p$^2$, 
4s4d, 
4s5d, 
4s6d, 
4s5s, 
4s6s, 
4p4f, 
4s4p, 
4s5p, 
4s6p, 
4s4f, 
4s5f, 
4s6f, 
4p5s, and 
4p4d 
configurations in \ion{Kr}{vii}. 

The numerical values of the parameters adopted in the present calculations are reported in 
Tables \ref{tab:kriv:para}-\ref{tab:krvii:para}, 
while the computed energies for \ion{Kr}{iv-vii} are compared with available experimental values in 
Tables \ref{tab:kriv:ener}-\ref{tab:krvii:ener}, 
respectively. 
Tables \ref{tab:kriv:loggf}-\ref{tab:krvii:loggf} 
give the computed oscillator strengths ($\log gf$) 
and transition probabilities ($gA$, in s$^{-1}$) for \ion{Kr}{iv-vii}, respectively, and the numerical values (in cm$^{-1}$) 
of lower and upper energy levels, together with the corresponding wavelengths (in \AA). The cancellation factor, 
CF, as defined by \citet{cowan1981} is also given in the last column of each table. For a specific transition, 
a very small value of this parameter (typically $< 0.05$) indicates strong cancellation effects in the 
calculation of the line strength. In this case, the corresponding oscillator strength and transition 
probability could be affected by larger uncertainties and, as a consequence, should be considered with  care. 

Tables \ref{tab:kriv:loggf}-\ref{tab:krvii:loggf} are provided in VO\footnote{Virtual Observatory}-compliant format
via the registered\footnote{cf., \url{http://dc.zah.uni-heidelberg.de/wirr/q/ui/fixed}} T\"ubingen Oscillator Strengths Service 
\citep[TOSS\footnote{\url{http://dc.g-vo.org/TOSS}},][]{rauchetal2016mo} that has recently been developed by GAVO.

\onltab{
\onecolumn
\begin{longtable}{llrrrl}
\caption{\label{tab:kriv:para}Radial parameters (in cm$^{-1}$) adopted for the calculations in \ion{Kr}{iv}.} \\
\hline\hline
\noalign{\smallskip}
Configuration & Parameter & HFR & Fitted & Ratio & Note$^a$ \\ 
\noalign{\smallskip}
\hline
\endfirsthead
\caption{continued.}\\
\hline\hline
\noalign{\smallskip}
Configuration & Parameter & HFR & Fitted & Ratio & Note$^a$ \\
\noalign{\smallskip}
\hline
\noalign{\smallskip}
\endhead
\hline
\noalign{\smallskip}
\endfoot
\noalign{\smallskip}
\multicolumn{6}{c}{Even parity} \\
4p$^3$           & E$_{av}$          &   31247 &  30541 &       &    \\
                 & F$^2$(4p,4p)      &   63340 &  56159 & 0.887 &    \\
                 & $\alpha$          &       0 & $-$151 &       &    \\
                 & $\zeta$$_{4p}$    &    4216 &   4645 & 1.102 &    \\
4p$^2$5p         & E$_{av}$          &  263829 & 253120 &       &    \\
                 & F$^2$(4p,4p)      &   65722 &  52042 & 0.792 &    \\
                 & $\alpha$          &       0 & $-$131 &       &    \\
\hline
\end{longtable}
$^a$ F: Fixed parameter value; R: ratios of these parameters had been fixed in the fitting process.\\
\noindent
\twocolumn
}

\onltab{
\onecolumn
\begin{longtable}{llrrrl}
\caption{\label{tab:krvv:para}Radial parameters (in cm$^{-1}$) adopted for the calculations in \ion{Kr}{v}.} \\
\hline\hline
\noalign{\smallskip}
Configuration & Parameter & HFR & Fitted & Ratio & Note$^a$ \\ 
\noalign{\smallskip}
\hline
\endfirsthead
\caption{continued.}\\
\hline\hline
\noalign{\smallskip}
Configuration & Parameter & HFR & Fitted & Ratio & Note$^a$ \\
\noalign{\smallskip}
\hline
\noalign{\smallskip}
\endhead
\hline
\noalign{\smallskip}
\endfoot
\noalign{\smallskip}
\multicolumn{6}{c}{Even parity} \\
4p$^2$           & E$_{av}$          &   23408 &  23750 &       &    \\
                 & F$^2$(4p,4p)      &   66066 &  57912 & 0.877 &    \\
                 & $\alpha$          &       0 &  $-$98 &       &    \\
                 & $\zeta$$_{4p}$    &    4645 &   5105 & 1.099 &    \\
4p5p             & E$_{av}$          &  300733 & 291773 &       &    \\
                 & $\zeta$$_{4p}$    &    5032 &   4888 & 0.971 &    \\
                 & $\zeta$$_{5p}$    &    1264 &   1182 & 0.935 &    \\
\hline
\end{longtable}
$^a$ F: Fixed parameter value.\\
\noindent
\twocolumn
}

\onltab{
\onecolumn
\begin{longtable}{llrrrl}
\caption{\label{tab:krvi:para}Radial parameters (in cm$^{-1}$) adopted for the calculations in \ion{Kr}{vi}.} \\
\hline\hline
\noalign{\smallskip}
Configuration & Parameter & HFR & Fitted & Ratio & Note$^a$ \\ 
\noalign{\smallskip}
\hline
\endfirsthead
\caption{continued.}\\
\hline\hline
\noalign{\smallskip}
Configuration & Parameter & HFR & Fitted & Ratio & Note$^a$ \\
\noalign{\smallskip}
\hline
\noalign{\smallskip}
\endhead
\hline
\noalign{\smallskip}
\endfoot
\noalign{\smallskip}
\multicolumn{6}{c}{Odd parity} \\
4p               & E$_{av}$          &   15431 &  15919 &       &    \\
                 & $\zeta$$_{4p}$    &    5091 &   5546 & 1.089 &    \\
5p               & E$_{av}$          &  337490 & 329220 &       &    \\
                 & $\zeta$$_{5p}$    &    1617 &   1767 & 1.093 &    \\
4s4p4d           & E$_{av}$          &  350565 & 351918 &       &    \\
                 & $\zeta$$_{4p}$    &    5220 &   5415 & 1.037 &    \\
                 & $\zeta$$_{4d}$    &     346 &    346 & 1.000 & F  \\
\hline
\end{longtable}
$^a$ F: Fixed parameter value.\\
\noindent
\twocolumn
}

\onltab{
\onecolumn
\begin{longtable}{llrrrl}
\caption{\label{tab:krvii:para}Radial parameters (in cm$^{-1}$) adopted for the calculations in \ion{Kr}{vii}.} \\
\hline\hline
\noalign{\smallskip}
Configuration & Parameter & HFR & Fitted & Ratio & Note$^a$ \\ 
\noalign{\smallskip}
\hline
\endfirsthead
\caption{continued.}\\
\hline\hline
\noalign{\smallskip}
Configuration & Parameter & HFR & Fitted & Ratio & Note$^a$ \\
\noalign{\smallskip}
\hline
\noalign{\smallskip}
\endhead
\hline
\noalign{\smallskip}
\endfoot
\noalign{\smallskip}
\multicolumn{6}{c}{Even parity} \\
4s$^2$           & E$_{av}$          &    8227 &   8218 &       &    \\
4p$^2$           & E$_{av}$          &  289178 & 291671 &       &    \\
                 & F$^2$(4p,4p)      &   70911 &  62322 & 0.879 &    \\
                 & $\alpha$          &       0 &    422 &       &    \\
                 & $\zeta$$_{4p}$    &    5539 &   6051 & 1.093 &    \\
4s4d             & E$_{av}$          &  353888 & 356691 &       &    \\
                 & $\zeta$$_{4d}$    &     409 &    465 & 1.139 &    \\
\hline
\end{longtable}
$^a$ F: Fixed parameter value.\\
\noindent
\twocolumn
}

\onltab{
\onecolumn
\begin{longtable}{rrrrl}
\caption{\label{tab:kriv:ener} Comparison between available experimental and calculated energy levels in \ion{Kr}{iv}. Energies are given in cm$^{-1}$.} \\
\hline\hline
\noalign{\smallskip}
E$_{exp}$$^a$  & E$_{calc}$$^b$ & $\Delta$E & J & Leading components (in \%) in $LS$ coupling$^c$ \\ 
\noalign{\smallskip}
\hline
\endfirsthead
\caption{continued.}\\
\hline\hline
\noalign{\smallskip}
E$_{exp}$$^a$  & E$_{calc}$$^b$ & $\Delta$E & J & Leading components (in \%) in $LS$ coupling$^c$ \\
\noalign{\smallskip}
\hline
\noalign{\smallskip}
\endhead
\hline
\noalign{\smallskip}
\endfoot
\noalign{\smallskip}
\multicolumn{5}{c}{Odd parity} \\
        0.0   &      0 &     0  & 1.5 &   95  4p$^3$ $^4$S  \\   
    17037.6   &  17022 &    16  & 1.5 &   87  4p$^3$ $^2$D  +   10  4p$^3$ $^2$P  \\  
    18700.3   &  18716 & $-$16  & 2.5 &   97  4p$^3$ $^2$D  \\   
    31056.4   &  31067 & $-$11  & 0.5 &   96  4p$^3$ $^2$P  \\   
    33405.6   &  33395 &    11  & 1.5 &   85  4p$^3$ $^2$P  +   10  4p$^3$ $^2$D  \\   
   234827.9   & 234724 &   104  & 0.5 &   64  4p$^2$($^3$P)5p $^2$S  +   19  4p$^2$($^3$P)5p $^4$D  +    7  4p$^2$($^3$P)5p $^4$P  \\   
   237445.1   & 237386 &    59  & 0.5 &   74  4p$^2$($^3$P)5p $^4$D  +   23  4p$^2$($^3$P)5p $^2$S  \\
\hline
\end{longtable}
\noindent
$^{(a)}$ From  \citet{saloman2007}.\\
$^{(b)}$ This work.\\
$^{(c)}$ Only the first three components that are larger than 5\% are given.
\twocolumn
}

\onltab{
\onecolumn
\begin{longtable}{rrrrl}
\caption{\label{tab:krv:ener} Comparison between available experimental and calculated energy levels in \ion{Kr}{v}. Energies are given in cm$^{-1}$.} \\
\hline\hline
\noalign{\smallskip}
E$_{exp}$$^a$  & E$_{calc}$$^b$ & $\Delta$E & J & Leading components (in \%) in $LS$ coupling$^c$ \\ 
\noalign{\smallskip}
\hline
\endfirsthead
\caption{continued.}\\
\hline\hline
\noalign{\smallskip}
E$_{exp}$$^a$  & E$_{calc}$$^b$ & $\Delta$E & J & Leading components (in \%) in $LS$ coupling$^c$ \\
\noalign{\smallskip}
\hline
\noalign{\smallskip}
\endhead
\hline
\noalign{\smallskip}
\endfoot
\noalign{\smallskip}
\multicolumn{5}{c}{Even parity} \\
        0.0  &   $-$33 &     33 &  0  &  94  4p$^2$ $^3$P  \\
     3742.8  &    3782 &  $-$39 &  1  &  97  4p$^2$ $^3$P  \\ 
     7595.3  &    7630 &  $-$34 &  2  &  88  4p$^2$ $^3$P  +   9  4p$^2$ $^1$D  \\ 
    19722.9  &   19722 &      1 &  2  &  88  4p$^2$ $^1$D  +   9  4p$^2$ $^3$P  \\ 
    39203.9  &   39086 &    118 &  0  &  93  4p$^2$ $^1$S  \\ 
   278654.0  &  276953 &   1701 &  2  &  48  4p$^4$ $^3$P  +  41  4s4p$^2$4d $^3$P  \\ 
   278928.0  &  279073 & $-$145 &  1  &  54  4p5p $^1$P    +  42  4p5p $^3$D  \\
\hline
\end{longtable}
\noindent
$^{(a)}$ From  \citet{saloman2007} and \citet{rezendeetal2010}.\\
$^{(b)}$ This work.\\
$^{(c)}$ Only the first three components that are larger than 5\% are given.
\twocolumn
}

\onltab{
\onecolumn
\begin{longtable}{rrrrl}
\caption{\label{tab:krvi:ener} Comparison between available experimental and calculated energy levels in \ion{Kr}{vi}. Energies are given in cm$^{-1}$.} \\
\hline\hline
\noalign{\smallskip}
E$_{exp}$$^a$  & E$_{calc}$$^b$ & $\Delta$E & J & Leading components (in \%) in $LS$ coupling$^c$ \\ 
\noalign{\smallskip}
\hline
\endfirsthead
\caption{continued.}\\
\hline\hline
\noalign{\smallskip}
E$_{exp}$$^a$  & E$_{calc}$$^b$ & $\Delta$E & J & Leading components (in \%) in $LS$ coupling$^c$ \\
\noalign{\smallskip}
\hline
\noalign{\smallskip}
\endhead
\hline
\noalign{\smallskip}
\endfoot
\noalign{\smallskip}
\multicolumn{5}{c}{Odd parity} \\
        0 &     $-$1 &      1 & 0.5 &   97  4p $^2$P  \\ 
     8110 &     8114 &   $-$4 & 1.5 &   97  4p $^2$P  \\  
   276011 &   275887 &    124 & 1.5 &   58  4p$^3$ $^2$D  +   20  4s4p($^3$P)4d $^2$D  +   15  4p$^3$ $^4$S \\  
   278062 &   277979 &     83 & 2.5 &   73  4p$^3$ $^2$D  +   25  4s4p($^3$P)4d $^2$D  \\  
   278787 &   278812 &  $-$25 & 1.5 &   81  4p$^3$ $^4$S  +   13  4p$^3$ $^2$D  \\  
   303697 &   303768 &  $-$71 & 0.5 &   81  4p$^3$ $^2$P  +   12  4s4p($^3$P)4d $^2$P  \\  
   305385 &   305317 &     68 & 1.5 &   73  4p$^3$ $^2$P  +   13  4s4p($^3$P)4d $^2$P  \\
\hline
\end{longtable}
\noindent
$^{(a)}$ From  \citet{saloman2007} and \citet{fariasetal2011}.\\
$^{(b)}$ This work.\\
$^{(c)}$ Only the first three components that are larger than 5\% are given.
\twocolumn
}

\onltab{
\onecolumn
\begin{longtable}{rrrrl}
\caption{\label{tab:krvii:ener} Comparison between available experimental and calculated energy levels in \ion{Kr}{vii}. Energies are given in cm$^{-1}$.} \\
\hline\hline
\noalign{\smallskip}
E$_{exp}$$^a$  & E$_{calc}$$^b$ & $\Delta$E & J & Leading components (in \%) in $LS$ coupling$^c$ \\ 
\noalign{\smallskip}
\hline
\endfirsthead
\caption{continued.}\\
\hline\hline
\noalign{\smallskip}
E$_{exp}$$^a$  & E$_{calc}$$^b$ & $\Delta$E & J & Leading components (in \%) in $LS$ coupling$^c$ \\
\noalign{\smallskip}
\hline
\noalign{\smallskip}
\endhead
\hline
\noalign{\smallskip}
\endfoot
\noalign{\smallskip}
\multicolumn{5}{c}{Even parity} \\
        0.0 &        0 &      0 & 0 &   98  4s$^2$ $^1$S  \\ 
   274931.7 &   274895 &     36 & 0 &   96  4p$^2$ $^3$P  \\  
   279414.5 &   279463 &  $-$49 & 1 &  100  4p$^2$ $^3$P  \\  
   279714.8 &   279726 &  $-$12 & 2 &   58  4p$^2$ $^1$D  +   32  4p$^2$ $^3$P  +   10  4s4d $^1$D \\  
   288190.2 &   288165 &     26 & 2 &   68  4p$^2$ $^3$P  +   27  4p$^2$ $^1$D  +    5  4s4d $^1$D \\  
   321794.0 &   321795 &   $-$1 & 0 &   93  4p$^2$ $^1$S  \\  
   349973.1 &   349973 &      0 & 1 &  100  4s4d $^3$D  \\
\hline
\end{longtable}
\noindent
$^{(a)}$ From  \citet{rainerietal2014}.\\
$^{(b)}$ This work.\\
$^{(c)}$ Only the first three components that are larger than 5\% are given.
\twocolumn
}

\onltab{
\onecolumn
\begin{longtable}{rrccrrccrrr}
\caption{\label{tab:kriv:loggf}Calculated HFR oscillator strengths ($\log gf$) and transition probabilities ($gA$) 
                               in \ion{Kr}{iv}. CF is absolute value of the cancellation factor as defined by \citet{cowan1981}. 
                               In cols\@. 3 and 6, e is written for even and o for odd.} \\
\hline\hline
\noalign{\smallskip}
                 & \multicolumn{3}{c}{Lower level} && \multicolumn{3}{c}{Upper level} &           &      &    \\
\cline{2-4}\cline{6-8}
\noalign{\smallskip}
Wavelength{$^a$} / \AA & Energy{$^b$} / cm$^{-1}$ & Parity & $J$  && Energy{$^b$} / cm$^{-1}$ & Parity & $J$  & $\log gf$ & $gA$ / s$^{-1}$  & CF \\
\noalign{\smallskip}
\hline
\endfirsthead
\caption{continued.}\\
\hline\hline
\noalign{\smallskip}
                 & \multicolumn{3}{c}{Lower level} && \multicolumn{3}{c}{Upper level} &           &      &    \\
\cline{2-4}\cline{6-8}
\noalign{\smallskip}
Wavelength{$^a$} / \AA & Energy{$^b$} / cm$^{-1}$ & Parity & $J$  && Energy{$^b$} / cm$^{-1}$ & Parity & $J$  & $\log gf$ & $gA$ / s$^{-1}$  & CF \\
\noalign{\smallskip}
\hline
\noalign{\smallskip}
\endhead
\hline
\noalign{\smallskip}
\endfoot
\noalign{\smallskip}
\\
\multicolumn{10}{p{10cm}}{Table only available via the German Astrophysical Virtual Observatory (GAVO) service TOSS (http://dc.g-vo.org/TOSS).}\\
\\
\end{longtable}
\twocolumn
}

\onltab{
\onecolumn
\begin{longtable}{rrccrrccrrr}
\caption{\label{tab:krv:loggf}Calculated HFR oscillator strengths ($\log gf$) and transition probabilities ($gA$) 
                               in \ion{Kr}{v}. CF is absolute value of the cancellation factor as defined by \citet{cowan1981}. 
                               In cols\@. 3 and 6, e is written for even and o for odd.} \\
\hline\hline
\noalign{\smallskip}
                 & \multicolumn{3}{c}{Lower level} && \multicolumn{3}{c}{Upper level} &           &      &    \\
\cline{2-4}\cline{6-8}
\noalign{\smallskip}
Wavelength{$^a$} / \AA & Energy{$^b$} / cm$^{-1}$ & Parity & $J$  && Energy{$^b$} / cm$^{-1}$ & Parity & $J$  & $\log gf$ & $gA$ / s$^{-1}$  & CF \\
\noalign{\smallskip}
\hline
\endfirsthead
\caption{continued.}\\
\hline\hline
\noalign{\smallskip}
                 & \multicolumn{3}{c}{Lower level} && \multicolumn{3}{c}{Upper level} &           &      &    \\
\cline{2-4}\cline{6-8}
\noalign{\smallskip}
Wavelength{$^a$} / \AA & Energy{$^b$} / cm$^{-1}$ & Parity & $J$  && Energy{$^b$} / cm$^{-1}$ & Parity & $J$  & $\log gf$ & $gA$ / s$^{-1}$  & CF \\
\noalign{\smallskip}
\hline
\noalign{\smallskip}
\endhead
\hline
\noalign{\smallskip}
\endfoot
\noalign{\smallskip}
\\
\multicolumn{10}{p{10.0cm}}{Table only available via the German Astrophysical Virtual Observatory (GAVO) service TOSS (http://dc.g-vo.org/TOSS).}\\
\\
\end{longtable}
\twocolumn
}

\onltab{
\onecolumn
\begin{longtable}{rrccrrccrrr}
\caption{\label{tab:krvi:loggf}Calculated HFR oscillator strengths ($\log gf$) and transition probabilities ($gA$) 
                               in \ion{Kr}{vi}. CF is absolute value of the cancellation factor as defined by \citet{cowan1981}. 
                               In cols\@. 3 and 6, e is written for even and o for odd.} \\
\hline\hline
\noalign{\smallskip}
                 & \multicolumn{3}{c}{Lower level} && \multicolumn{3}{c}{Upper level} &           &      &    \\
\cline{2-4}\cline{6-8}
\noalign{\smallskip}
Wavelength{$^a$} / \AA & Energy{$^b$} / cm$^{-1}$ & Parity & $J$  && Energy{$^b$} / cm$^{-1}$ & Parity & $J$  & $\log gf$ & $gA$ / s$^{-1}$  & CF \\
\noalign{\smallskip}
\hline
\endfirsthead
\caption{continued.}\\
\hline\hline
\noalign{\smallskip}
                 & \multicolumn{3}{c}{Lower level} && \multicolumn{3}{c}{Upper level} &           &      &    \\
\cline{2-4}\cline{6-8}
\noalign{\smallskip}
Wavelength{$^a$} / \AA & Energy{$^b$} / cm$^{-1}$ & Parity & $J$  && Energy{$^b$} / cm$^{-1}$ & Parity & $J$  & $\log gf$ & $gA$ / s$^{-1}$  & CF \\
\noalign{\smallskip}
\hline
\noalign{\smallskip}
\endhead
\hline
\noalign{\smallskip}
\endfoot
\noalign{\smallskip}
\\
\multicolumn{10}{p{10.0cm}}{Table only available via the German Astrophysical Virtual Observatory (GAVO) service TOSS (http://dc.g-vo.org/TOSS).}\\
\\
\end{longtable}
\twocolumn
}

\onltab{
\onecolumn
\begin{longtable}{rrccrrccrrr}
\caption{\label{tab:krvii:loggf}Calculated HFR oscillator strengths ($\log gf$) and transition probabilities ($gA$) 
                               in \ion{Kr}{vii}. CF is absolute value of the cancellation factor as defined by \citet{cowan1981}. 
                               In cols\@. 3 and 6, e is written for even and o for odd.} \\
\hline\hline
\noalign{\smallskip}
                 & \multicolumn{3}{c}{Lower level} && \multicolumn{3}{c}{Upper level} &           &      &    \\
\cline{2-4}\cline{6-8}
\noalign{\smallskip}
Wavelength{$^a$} / \AA & Energy{$^b$} / cm$^{-1}$ & Parity & $J$  && Energy{$^b$} / cm$^{-1}$ & Parity & $J$  & $\log gf$ & $gA$ / s$^{-1}$  & CF \\
\noalign{\smallskip}
\hline
\endfirsthead
\caption{continued.}\\
\hline\hline
\noalign{\smallskip}
                 & \multicolumn{3}{c}{Lower level} && \multicolumn{3}{c}{Upper level} &           &      &    \\
\cline{2-4}\cline{6-8}
\noalign{\smallskip}
Wavelength{$^a$} / \AA & Energy{$^b$} / cm$^{-1}$ & Parity & $J$  && Energy{$^b$} / cm$^{-1}$ & Parity & $J$  & $\log gf$ & $gA$ / s$^{-1}$  & CF \\
\noalign{\smallskip}
\hline
\noalign{\smallskip}
\endhead
\hline
\noalign{\smallskip}
\endfoot
\noalign{\smallskip}
\\
\multicolumn{10}{p{10.0cm}}{Table only available via the German Astrophysical Virtual Observatory (GAVO) service TOSS (http://dc.g-vo.org/TOSS).}\\
\\
\end{longtable}
\twocolumn
}

\begin{figure}
   \resizebox{\hsize}{!}{\includegraphics{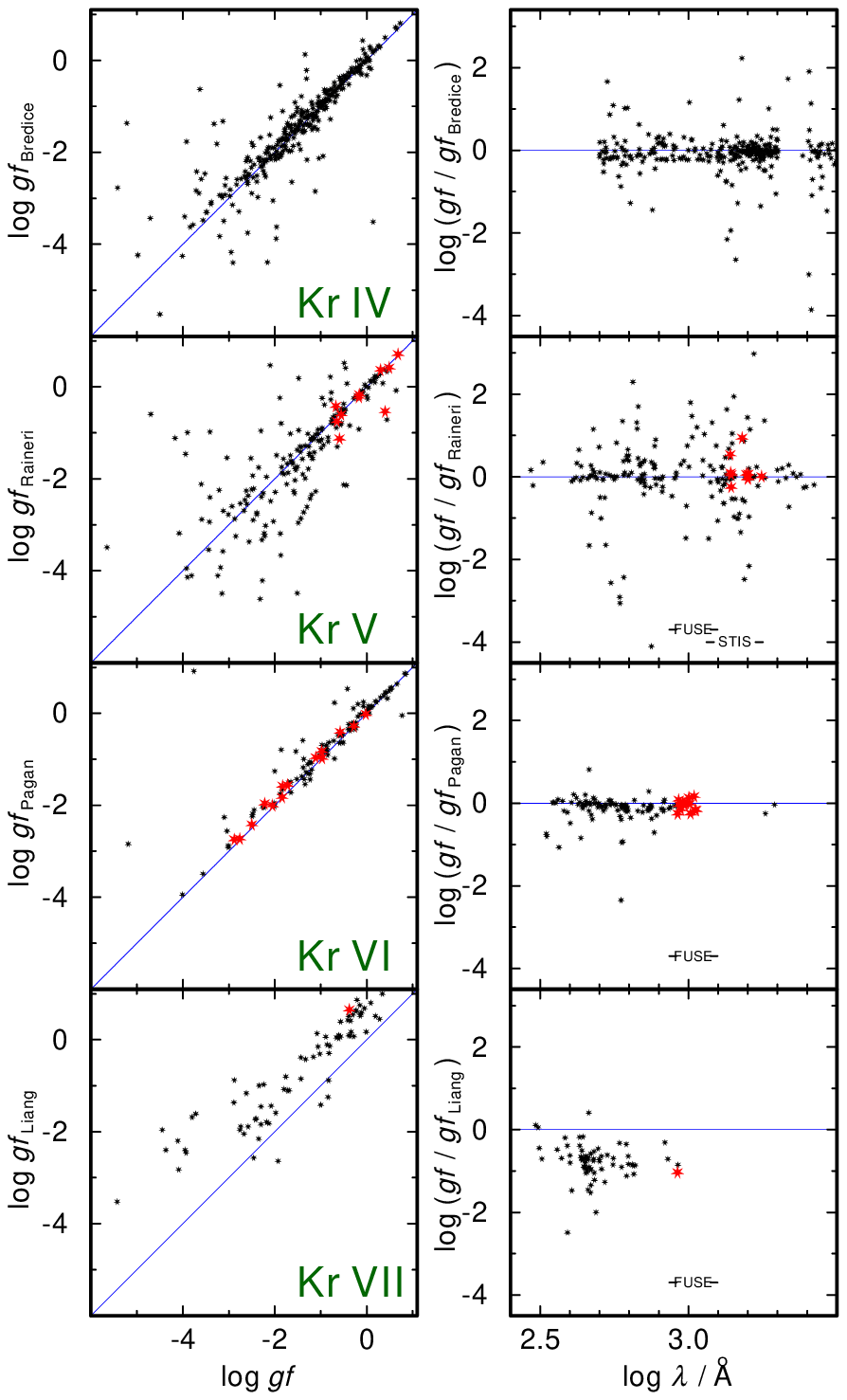}}
    \caption{Comparison of our weighted oscillator strengths for \ion{Kr}{iv-vii} (from top to
             bottom) to those of
             \citet{brediceetal2000},
             \citet{rainerietal2012},
             \citet{paganetal1996}, and
             \citet{liangetal2013}, respectively.
             Left panel: Comparison of weighted oscillator strengths.
             Right panel: Ratio of weighted oscillator strengths over wavelength.
             The wavelength ranges of our FUSE and HST/STIS spectra are marked.
             The larger, red symbols refer to the lines identified in \re (see Figs.\,\ref{fig:krwerner} and \ref{fig:krnew}).
           }
   \label{fig:fcomp}
\end{figure}

For \ion{Kr}{iv-vii}, oscillator strengths were previously published by several authors. In the following,
we compare our new data to theirs. Figure\,\ref{fig:fcomp} illustrates this comparison.

\citet{brediceetal2000} calculated weighted oscillator strengths for 471 spectral lines of \ion{Kr}{iv}
involving the 
4s$^2$4p$^3$, 
4s4p$^4$, 
4s$^2$4p$^2$5p, and 
4s$^2$4p$^2$(5s + 6s + 4d + 5d)
configurations within the wavelength interval [501.50\,\AA, 4703.85\,\AA]. Of these lines, 320 agree
within 0.1\,\AA\ to lines reported in  Table\,\ref{tab:kriv:loggf}. We selected them to compare their 
$\log gf$ values to ours in Fig.\ref{fig:fcomp}. As seen from this figure, even if both sets of results 
agree for many lines, a large scatter is also observed for many other lines. This is obviously due to 
the very limited multiconfiguration Hartree-Fock model (including only two odd- and seven even-parity 
configurations) that were considered by \citet{brediceetal2000}.

\citet{rainerietal2012} published transition probabilities for 313 lines of \ion{Kr}{v}
([294.27\,\AA, 3614.10\,\AA]).
They were calculated with Cowan's package \citep{cowan1981}, i.e., the Hartree-Fock method with relativistic corrections
using energy parameters from least-squares and dipole-reduced matrix from a core polarization calculation. A large set
of 
4s$^2$4p$^2$, 
4p5p, 
4p$^4$, 
4s4p$^2$4d, 
4p4f, 
4s4p$^2$5s,
4s$^2$4d$^2$, 
4s4p4d5p, 
4p$^3$4f, 
4s4p4d4f, 
4s$^2$4f$^2$, and 
4s4d$^3$ even
and 
4s4p$^3$, 
4p4d, 
4p5s, 
4p5d, 
4p6s, 
4p5g, 
4p6d, 
4s4p$^2$5p,
4s4p$^2$4f, 
4p$^3$4d, 
4s4p$^2$6p, 
4s4p4d$^2$, 
4p$^3$5s, 
4s$^2$4d4f,
4s$^2$4f5s, 
4p$^3$5d, 
4s$^2$4f5d, 
4s4d$^2$5p, and 
4s4d$^2$4f 
odd configurations was considered.

We selected 183 of these lines (by wavelength agreement within 0.1\,\AA\ to lines in Table\,\ref{tab:krv:loggf}) 
and compared their $\log gf$ values to ours (Fig.\ref{fig:fcomp}). Although a good agreement between the two sets 
of data is observed for many lines, a large scatter is also obtained for a number of transitions, in particular 
for those corresponding to weak oscillator strengths, i.e., $\log gf < -2$. This is mainly due to the rather large 
cancellation effects that appear in the calculations of these types of transition rates in both works. Moreover, it is worth 
noting that \citet{rainerietal2012} modified only their electric dipole matrix elements to take core-polarization 
effects into account while, in our work, all the radial wave functions were also modified by a model potential, 
including one- and two-body core-polarization contributions, together with a core-penetration correction 
\citep[see, e.g.,][]{quinetetal2002}. This could also explain some of the differences between the two sets of results.

\citet{paganetal1996} calculated 138 weighted \ion{Kr}{vi} oscillator strength  ([331.65\,\AA, 2051.72\,\AA]) in a multiconfigurational HFR approach
considering
4s$^2$4p, 
4p$^3$, 
4s$^2$5p, 
4s4p4d, 
4s$^2$4f, 
4s4p5s, 
4s$^2$6p, 
4s4d4f, 
4p$^2$4f, 
4p4d$^2$, and 
4d$^2$4f configurations for odd parity and 
4s4p$^2$, 
4s$^2$4d, 
4s$^2$5s, 
4s$^2$5d, 
4s4p4f, 
4s4p5s, 
4p$^2$4d, and
4s4d$^2$ for even parity.
Analogously to \ion{Kr}{v}, we compared the $\log gf$ values of 115 selected \ion{Kr}{vi} lines with our data (Fig.\ref{fig:fcomp}).
Although both sets of data are in good agreement, we note that our oscillator strengths are generally smaller than those obtained 
by \citet{paganetal1996}. This is essentially due to the much more extended multiconfiguration expansions and the core-polarization 
effects that we include in our calculations.

\citet{liangetal2013} presented oscillator strengths for 90 lines ([201.05\,\AA, 920.98\,\AA]) For \ion{Kr}{vii}.
For their calculations with the AUTOSTRUCTURE code \citep{badnell2011}, they only use nine configurations,
i.e., 
4s$^2$, 
4p$^2$, 
4s4d, 
4s5s, 
4s5d, and 
4s4p, 
4s4f, 
4p4d, 
4s5p 
for the even and odd parities, respectively. In particular, they omitt some configurations, such as 
4d$^2$ and 
4d4f which, according to our calculations,  appeared to have non-negligible interactions with 
4s$^2$, 
4p$^2$ and 
4s4p, 
4s4f, 
4p4d, respectively. The effect of this rather limited model is illustrated
in Fig.\ref{fig:fcomp}, where our $\log gf$ values, obtained with an extended configuration interaction approach,
are systematically smaller than those reported by \citet{liangetal2013}.

\section{Effective temperature and surface gravity}
\label{sect:tefflogg}

\citet{dreizlerwerner1996} analyzed the optical TWIN spectra (Sect.\,\ref{sect:observation}) with
NLTE model atmospheres that considered opacities of H, He, C, N, O, and Si. They derived
\Teffw{70\,000 \pm 5000} and $\log (g \mathrm{/ cm/s^2}) = 7.5 \pm 0.3$. 
In their spectral analysis based on H+He composed LTE model atmospheres,
\citet{vennesetal1998} determined \Teffw{68\,600 \pm 1800} and \loggw{7.20 \pm 0.07}.
\citet{dreizler1999} improved the metal abundance analysis of C, N, O, and Ni based on
HST/GHRS (Goddard High-Resolution Spectrograph) observations and used the previously determined
\Teffw{70\,000} and \loggw{7.5}. The same did \citet{werneretal2012} for their Kr and Xe abundance analysis.
In our latest models, much more species (Table\,\ref{tab:abre}) and, thus, a higher metal opacity 
is considered. Therefore, we start here with a new assessment of \Teff and \logg.

\begin{table}\centering 
  \caption{Photospheric abundances of \re.  
           IG is a generic model atom \citep{rauchdeetjen2003} comprising Ca, Sc, Ti, V, Cr, Mn, and Co.
           [X] denotes log (fraction\,/\,solar fraction) of species X.}
\label{tab:abre}
\setlength{\tabcolsep}{.4em}
\begin{tabular}{lr@{.}lr@{.}l}
\hline
\hline
\noalign{\smallskip}                                                                                          
                         & \multicolumn{4}{c}{Mass fraction}                                   \vspace{-2mm} \\
Element                  & \multicolumn{4}{c}{}                                                \vspace{-2mm} \\
\cline{2-5}
\noalign{ \smallskip}
                         & \multicolumn{2}{c}{Our models}  &\multicolumn{2}{c}{\citet{werneretal2012}} \\
\hline                   
\noalign{\smallskip}                                                                                   
\mmspr He    & $ 9$&$73\times 10^{-1}$ & $ \hbox{~}\hspace{5mm}9$&$78\times 10^{-1}$ \\  
\mmspr C     & $ 2$&$22\times 10^{-2}$ & $ 2$&$00\times 10^{-2}$ \\
\mmspr N     & $ 5$&$49\times 10^{-5}$ & $ 5$&$50\times 10^{-5}$ \\
\mmspr O     & $ 2$&$94\times 10^{-3}$ & $ 2$&$00\times 10^{-3}$ \\
\mmspr Si    & $ 1$&$60\times 10^{-4}$ & \multicolumn{2}{c}{~}  \\
\mmspr P     & $ 1$&$06\times 10^{-6}$ & \multicolumn{2}{c}{~}  \\
\mmspr S     & $ 3$&$96\times 10^{-5}$ & \multicolumn{2}{c}{~}  \\
\mmspr IG    & $ 9$&$98\times 10^{-7}$ & \multicolumn{2}{c}{~}  \\
\mmspr Fe    & $ 1$&$30\times 10^{-5}$ & \multicolumn{2}{c}{~}  \\
\mmspr Ni    & $ 7$&$25\times 10^{-5}$ & \multicolumn{2}{c}{~}  \\
\mmspr Zn    & $ 1$&$13\times 10^{-4}$ & \multicolumn{2}{c}{~}  \\
\mmspr Ga    & $ 3$&$44\times 10^{-5}$ & \multicolumn{2}{c}{~}  \\
\mmspr Ge    & $ 1$&$58\times 10^{-4}$ & \multicolumn{2}{c}{~}  \\
\mmspr As    & $ 1$&$60\times 10^{-5}$ & \multicolumn{2}{c}{~}  \\
\mmspr Kr    & $ 5$&$04\times 10^{-4}$ & $ 6$&$00\times 10^{-5}$ \\
\mmspr Mo    & $ 1$&$88\times 10^{-4}$ & \multicolumn{2}{c}{~}  \\
\mmspr Sn    & $ 2$&$04\times 10^{-4}$ & \multicolumn{2}{c}{~}  \\
\mmspr Xe    & $ 6$&$29\times 10^{-5}$ & $ 6$&$00\times 10^{-5}$ \\
\mmspr Ba    & $ 3$&$57\times 10^{-4}$ & \multicolumn{2}{c}{~}  \\
\hline
\end{tabular}
\end{table}

\begin{figure*}
   \resizebox{\hsize}{!}{\includegraphics{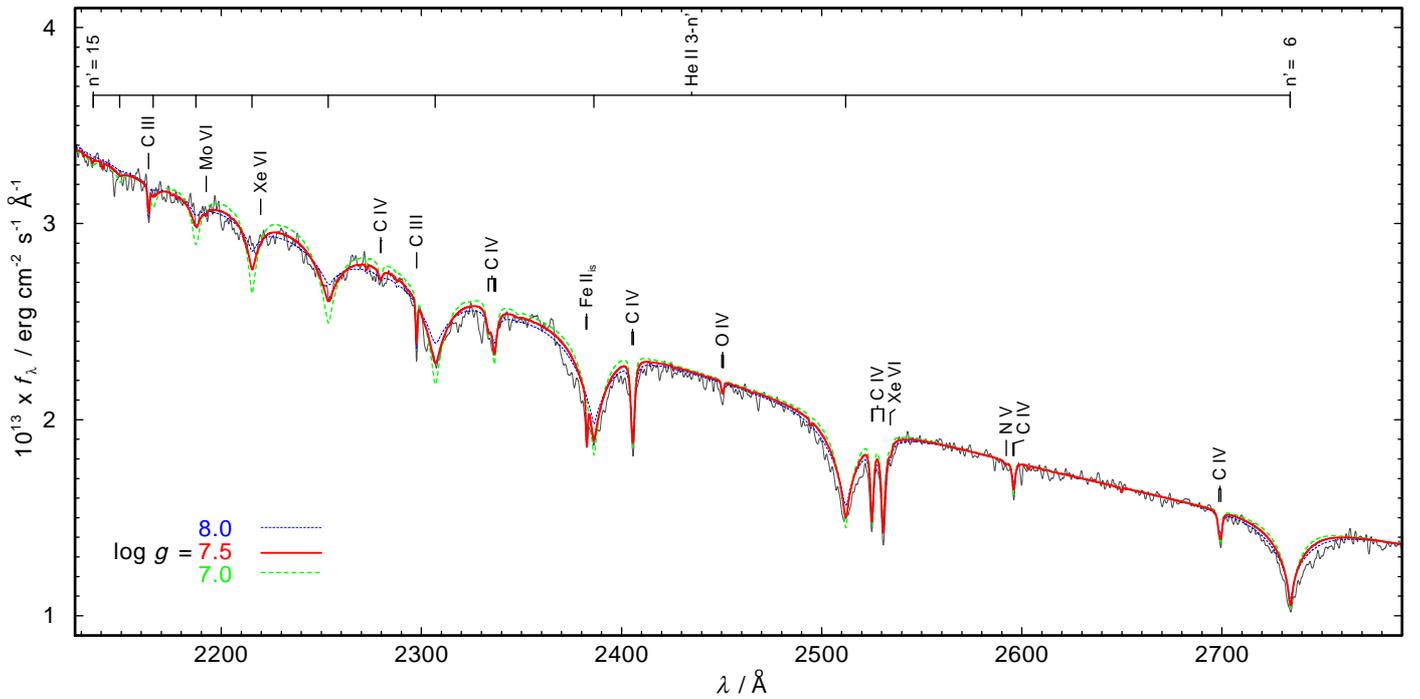}}
    \caption{Section of the HST/STIS spectrum, compared with models with different
             \logg of $7.0$ (thin, green, dashed), $7.5$ (thick, red), and $8.0$ (thin, blue, dashed)
             and \Teffw{70\,000}.
             The synthetic spectra are normalized to match the flux of the models
             at 2800\,\AA, respectively. 
             All spectra are convolved with Gaussians (full width at half maximum of 1\,\AA) for clarity. 
             Identified lines are indicated. ``is'' denotes interstellar origin. 
            }
   \label{fig:fowler}
\end{figure*}

The decrements of spectral series are very sensitive indicators for \logg
\citep[e.g.,][]{rauchetal1998,ziegleretal2012}. Figure\,\ref{fig:fowler} shows a comparison of
theoretical line profiles of the \ion{He}{ii} Fowler series (principal quantum numbers $n - n'$
with $n = 3$ and $n' \ge 4$) that are located in our HST/STIS spectrum. The central depressions
of \ion{He}{ii} $ 3 - [5, \ldots, 13]$ are well matched at \loggw{7.5}, while the decrement
is much too strong at \loggw{8.0,} and much too weak at \loggw{7.0}. We 
therefore verify \loggw{7.5}, which was the result of \citet{dreizlerwerner1996}, and
improved the error limit to $\pm 0.1\,\mathrm{dex}$.

\begin{figure*}
   \resizebox{\hsize}{!}{\includegraphics{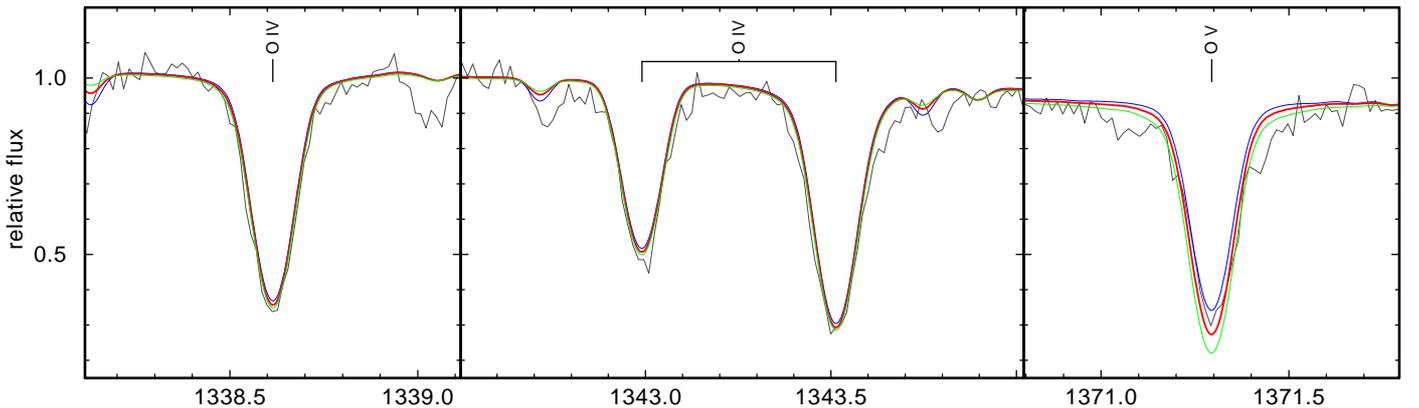}}
    \caption{Sections of the HST/STIS spectrum compared with three \loggw{7.5} models with different
             \Teff of
             66\,000\,K (thin, blue), 70\,000\,K (thick, red), and 74\,000\,K (thin, green).
             The synthetic spectra are normalized to match the observed flux at 1335\,\AA. 
            }
   \label{fig:Teff_O}
\end{figure*}

Figure\,\ref{fig:logg} demonstrates that the line wings 
of the \ion{He}{ii}
$n = 2$ and $n' \ge 3$ 
$n = 4$ and $n' \ge 5$ (Pickering) series 
are well in agreement with the observation at \loggw{7.5}. 
A different \logg cannot be compensated for by adapted interstellar \ion{H}{i} densities because this has
a strong impact on the inner line wings (Sect.\,\ref{sect:abundance}). The error range for these densities is below 20\%. 
Our optical spectra corroborate  the \logg determination. Figure\,\ref{fig:logg} shows the comparison of our
models to the observations. 
\ionw{He}{ii}{6560} is too shallow in all our models and does not match the TWIN observation. The reason is still unknown. 
The UVES spectra show a slightly better fit of this line, which may be a hint for some data-reduction
uncertainty in the TWIN and/or UVES spectra.
To make a comparison with the FUSE observation, we selected those \ion{He}{ii} $n=2$\,-\,n' lines
that are not contaminated by interstellar \ion{H}{i} line absorption.
Their observed line profiles and the series' decrement are well reproduced at \loggw{7.5}.
The insufficient blaze correction of the HST/STIS spectrum  only allows for an
evaluation of the inner line wings of \ionw{He}{ii}{1640.42} ($n=2-3$). 
\ionw{He}{ii}{1215.12} ($n=2-4$) is shown in Fig.\,\ref{fig:stisism}. 

\citet{dreizlerwerner1996} used the \ion{O}{iv}\,/\,\ion{O}{v} ionization equilibrium as an
indicator for \Teff. They found a simultaneous match of theoretical line profiles of
\ionww{O}{iv}{1338.6, 1343.0, 1343.5} (2p$^2$ $^2$P - 2p$^3$ $^2$D$^\mathrm{o}$) and
\ionw{O}{v}{1371.3} (2p $^1$P$^\mathrm{o}$ - 2p$^2$ $^1$D) to a high-resolution IUE observation.
In Fig.\,\ref{fig:Teff_O}, we show the same lines compared to our much better HST/STIS observation.
\ionw{O}{v}{1371.3} appears much more sensitive to \Teff, compared to the \ion{O}{iv} lines, and
\Teffw{70\,000} is verified within an error range of 2000\,K.

We adopt \Teffw{70\,000 \pm 2000} and \loggw{7.5 \pm 0.1}.
Many additional ionization equilibria, e.g., of
\ion{He}{i - ii} (Fig.\,\ref{fig:logg}),
\ion{C}{iii - iv} (Fig.\,\ref{fig:fowler}),
\ion{O}{iv - v} \citep[][and this paper, Fig.\,\ref{fig:Teff_O}]{dreizlerwerner1996},
\ion{Zn}{iv - v} \citep{rauchetal2014zn},
\ion{Ga}{iv - v} \citep{rauchetal2015ga},
\ion{Ge}{iv - vi} \citep{rauchetal2012ge},
\ion{Kr}{v - vii} \citep[][and this paper]{werneretal2012},
\ion{Mo}{v - vi} \citep{rauchetal2016mo},
\ion{Xe}{vi - vii} \citep{werneretal2012}, and
\ion{Ba}{vi - vii} \citep{rauchetal2014ba}
are well matched at these values.

\begin{landscape}
\addtolength{\textwidth}{6.3cm} 
\addtolength{\evensidemargin}{1cm}
\addtolength{\oddsidemargin}{1cm}
\begin{figure*}
\includegraphics[trim=0 -0 0 0,height=24.5cm,angle=270]{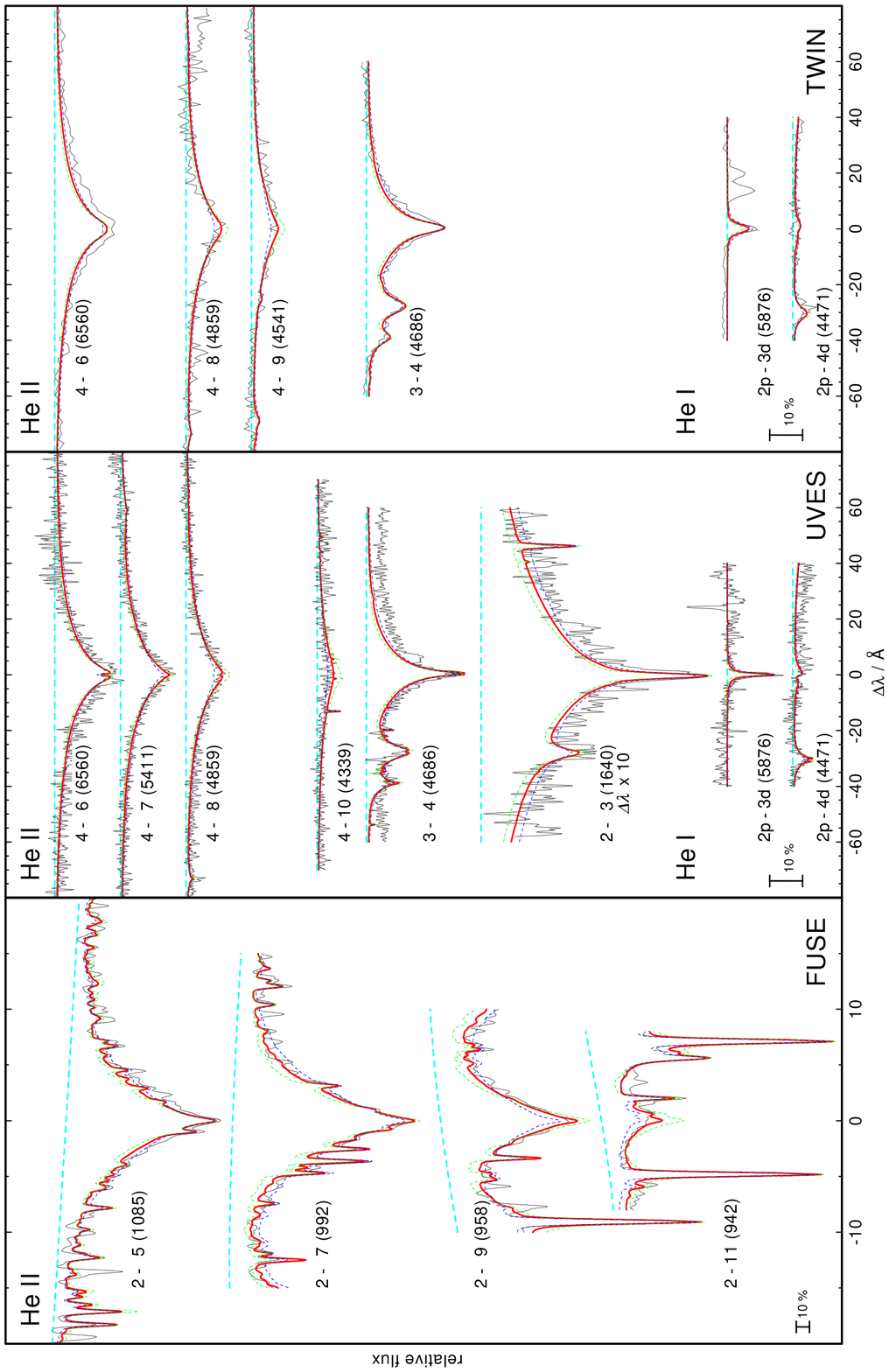}
  \caption{Three models (\Teffw{70\,000}) with 
           \logg = 7.0 (green, dashed), 7.5 (red), and 8.0 (blue, dashed)
           compared to 
           FUSE (left panel),
           UVES (middle), and
           TWIN (right) observations.
           The dashed, horizontal lines indicate the location of the local continuum. 
          }
  \label{fig:logg}
\end{figure*}
\end{landscape}

\section{Line identification and abundance analysis}
\label{sect:abundance}

\begin{figure}
   \resizebox{\hsize}{!}{\includegraphics{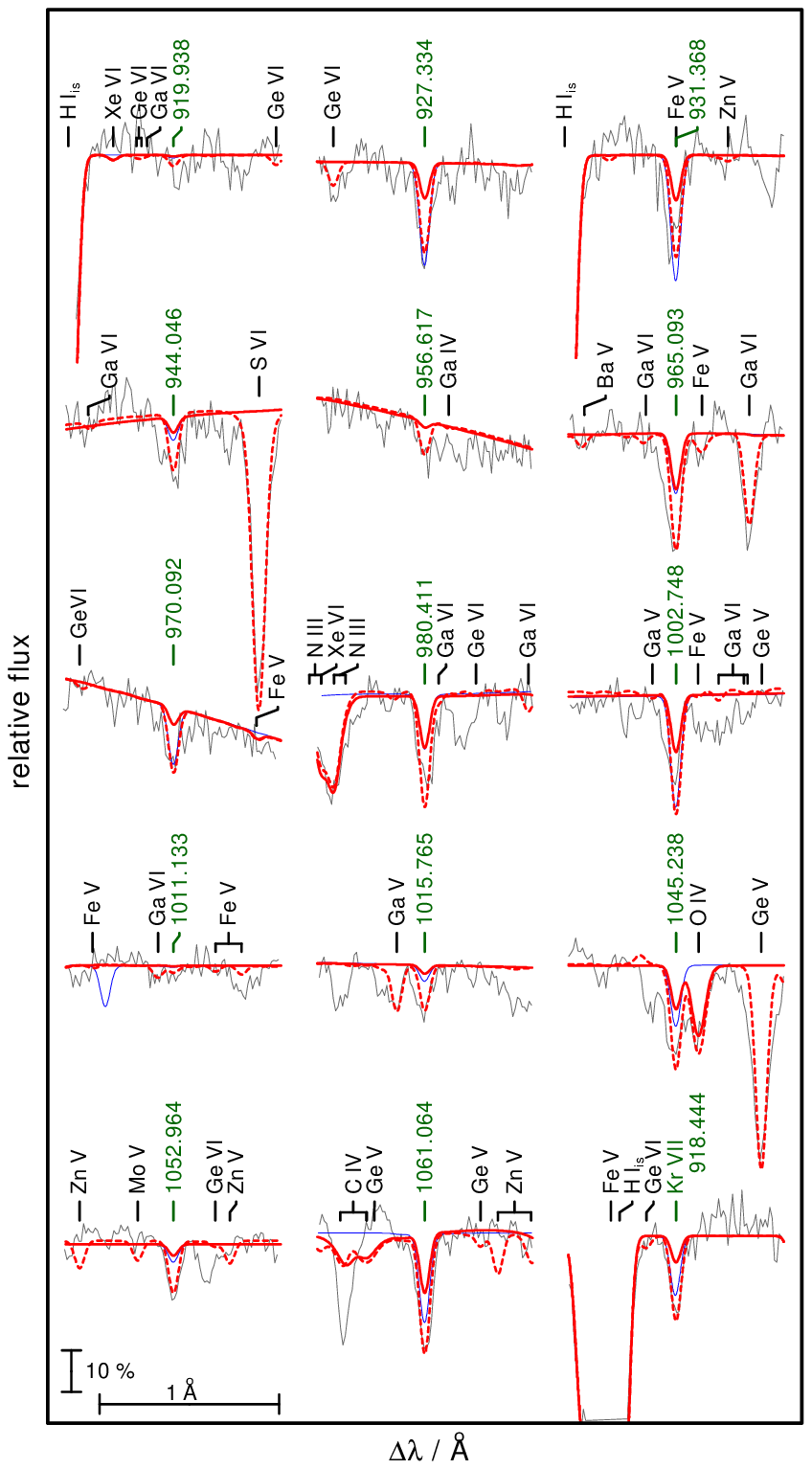}}
    \caption{\ion{Kr}{vi} lines and one \ion{Kr}{vii} line (bottom right) 
             identified by \citet{werneretal2012} in the observed FUSE spectrum of \re.
             Three synthetic spectra are overplotted.
             Thin, blue: the genuine spectrum of \citet[][\Teffw{70\,000}]{werneretal2012}, classical Kr model atom with $\log \mathrm{Kr} = -4.3$.
             Thick red (\Teffw{70\,000}): our new Kr model atom with $\log \mathrm{Kr} = -4.3$.
             Dashed, red (\Teffw{70\,000}): our new Kr model atom with $\log \mathrm{Kr} = -3.3$.
             Kr lines are indicated with their wavelengths from Tables \ref{tab:krvi:loggf} and
             \ref{tab:krvii:loggf}, and other lines by their ion's name. ``is'' denotes interstellar origin.
            }
   \label{fig:krwerner}
\end{figure}

We replaced our previously used Kr model atom and recalculated our latest model atmosphere \citep[see][]{rauchetal2016mo}. 
Figure\,\ref{fig:krwerner} shows that, in general, the wavelengths of the old and the new data are in 
good agreement, while the line strengths calculated with the new data are much smaller. 
There are three reasons for this deviation. First, the \ion{Kr}{iv-vii} model ions are much more
complete, e.g., for \ion{Kr}{vi} and \ion{Kr}{vii}, \citet{werneretal2012} constructed model ions
with 46 and 14 atomic levels that were combined with 140 and 2 line transitions with known oscillator strengths, 
respectively. These numbers were increased to 69 and 70 levels with 843 and 743 line transitions,
respectively (Table\,\ref{tab:ironic}). Second, the new oscillator strengths of the lines
that were used in the Kr abundance analysis are smaller in general (Table\,\ref{tab:oszi}).
Third (with minor impact), the chemical composition of the model atmospheres is different and 
the C, N, and O abundances were fine-tuned, i.e., the background opacity was increased and, thus,
the calculated atmospheric structure is different.

\begin{figure}
   \resizebox{\hsize}{!}{\includegraphics{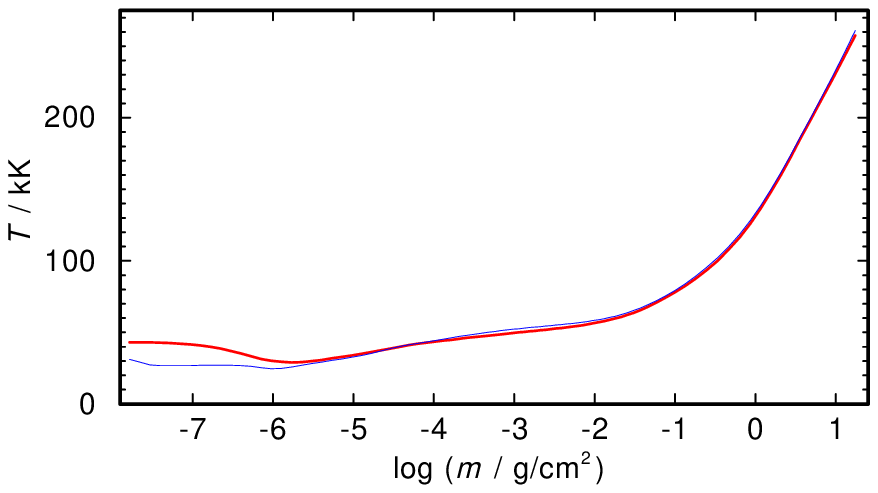}}
    \caption{Temperature structure of our model (thick, red) compared with the model of \citet[][thin, blue]{werneretal2012}.
            }
   \label{fig:temperature}
\end{figure}

In their Kr abundance analysis, \citet{werneretal2012} use He+C+N+O+Kr+Xe models and adopt the C, N, and O abundances
of \citet{dreizler1999}.
Our models also consider the opacities of Si, P, S, Ca, Sc, Ti, V, Cr, Mn, Fe, Co, Ni, Zn, Ga, Ge, As, Mo, Sn, and Ba. 
The C and O abundances were increased to better reproduce their observed lines.
The abundances are compared in Table\,\ref{tab:abre}. Figure\,\ref{fig:temperature} shows the temperature structures
of the respective models. Deviations are obvious in the outer atmosphere, but also in the line-forming
region ($-4 \la \log\,m \la +0.5$, $m$ is the column mass, measured from the outer boundary of our model atmospheres).

\begin{table}\centering 
  \caption{Comparison of $\log gf$ values of Kr lines used in the abundance analysis 
           (Figs.\,\ref{fig:krwerner}, \ref{fig:krnew}).}
\label{tab:oszi}
\setlength{\tabcolsep}{.4em}
\begin{tabular}{r@{.}lr@{.}lr@{.}l}
\hline
\hline
\noalign{\smallskip}                                                                                          
\multicolumn{2}{c}{~}                  & \multicolumn{4}{c}{$\log gf$}                              \vspace{-2mm} \\
\multicolumn{2}{c}{Wavelength\,/\,\AA} & \multicolumn{4}{c}{}                                       \vspace{-2mm} \\
\cline{3-6}
\noalign{ \smallskip}
\multicolumn{2}{c}{~}                  & \multicolumn{2}{c}{Literature} & \multicolumn{2}{c}{Our work} \\
\hline                   
\noalign{ \smallskip}
\multicolumn{6}{l}{\ion{Kr}{v}}                                                                        \\
1384&611                               & $-0$&590\tablefootmark{a}      & $-1$&128                     \\
1387&961                               & $-0$&650\tablefootmark{a}      & $-0$&749                     \\
1392&594                               & $-0$&670\tablefootmark{a}      & $-0$&420                     \\
1393&603                               & $-0$&170\tablefootmark{a}      & $-0$&224                     \\
1515&611                               & $ 0$&400\tablefootmark{a}      & $-0$&540                     \\
1566&073                               & $ 0$&300\tablefootmark{a}      & $ 0$&365                     \\
1583&456                               & $-0$&555\tablefootmark{a}      & $-0$&603                     \\
1589&269                               & $ 0$&680\tablefootmark{a}      & $ 0$&709                     \\
1591&875                               & $ 0$&490\tablefootmark{a}      & $ 0$&415                     \\
1764&478                               & $-0$&160\tablefootmark{a}      & $-0$&169                     \\
\noalign{ \smallskip}
\multicolumn{6}{l}{\ion{Kr}{vi}}                                                                       \\
 919&938                               & $-1$&950\tablefootmark{b}      & $-2$&22                      \\
 927&334                               & $-2$&420\tablefootmark{b}      & $-2$&50                      \\
 931&368                               & $-2$&005\tablefootmark{b}      & $-2$&05                      \\
 944&046                               & $-0$&959\tablefootmark{b}      & $-1$&11                      \\
 956&617                               & $-0$&974\tablefootmark{b}      & $-0$&97                      \\
 965&093                               & $-0$&019\tablefootmark{b}      & $-0$&02                      \\
 970&092                               & $-2$&731\tablefootmark{b}      & $-2$&76                      \\
 980&411                               & $-0$&278\tablefootmark{b}      & $-0$&28                      \\
1002&748                               & $-2$&748\tablefootmark{b}      & $-2$&88                      \\
1011&133                               & $-1$&826\tablefootmark{b}      & $-1$&84                      \\
1015&765                               & $-1$&580\tablefootmark{b}      & $-1$&83                      \\
1045&238                               & $-0$&823\tablefootmark{b}      & $-0$&98                      \\
1052&964                               & $-1$&561\tablefootmark{b}      & $-1$&72                      \\
1061&064                               & $-0$&409\tablefootmark{b}      & $-0$&58                      \\
\noalign{ \smallskip}
\multicolumn{6}{l}{\ion{Kr}{vii}}                                                                      \\
 918&444                               & $-0$&134\tablefootmark{c}      & $-0$&38                      \\
\noalign{\smallskip}                                                                                  
\hline
\end{tabular}
\tablefoot{
\tablefoottext{a}{\citet{rainerietal2012}}
\tablefoottext{b}{\citet{paganetal1996}}
\tablefoottext{c}{\citet{victortaylor1983}}
}
\end{table}

Due to the higher background opacity in our models, 
the calculated Kr lines are weaker compared with \citet[][$\log \mathrm{Kr} = -4.3 \pm 0.5$]{werneretal2012}. 
To match the observation, we have to increase the previously determined Kr mass fraction 
\citep[$\log \mathrm{Kr} = -4.3 \pm 0.5$,][]{werneretal2012}
by a factor of 10 to $5.1 \times 10 ^{-4}$ ($\log \mathrm{Kr} = -3.3 \pm 0.3$).
Our given error is estimated, considering the error propagation that is due to the uncertainties
of \Teff, \logg and the background opacity (Sect.\,\ref{sect:tefflogg}). 

\begin{figure}
   \resizebox{\hsize}{!}{\includegraphics{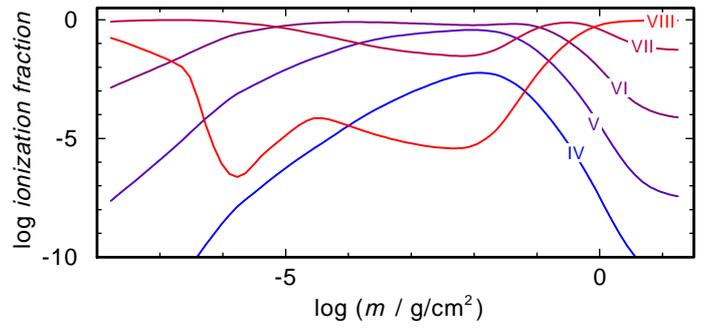}}
    \caption{Kr ionization fractions in our model for \re.
            }
   \label{fig:ion}
\end{figure}

\begin{figure}
   \resizebox{\hsize}{!}{\includegraphics{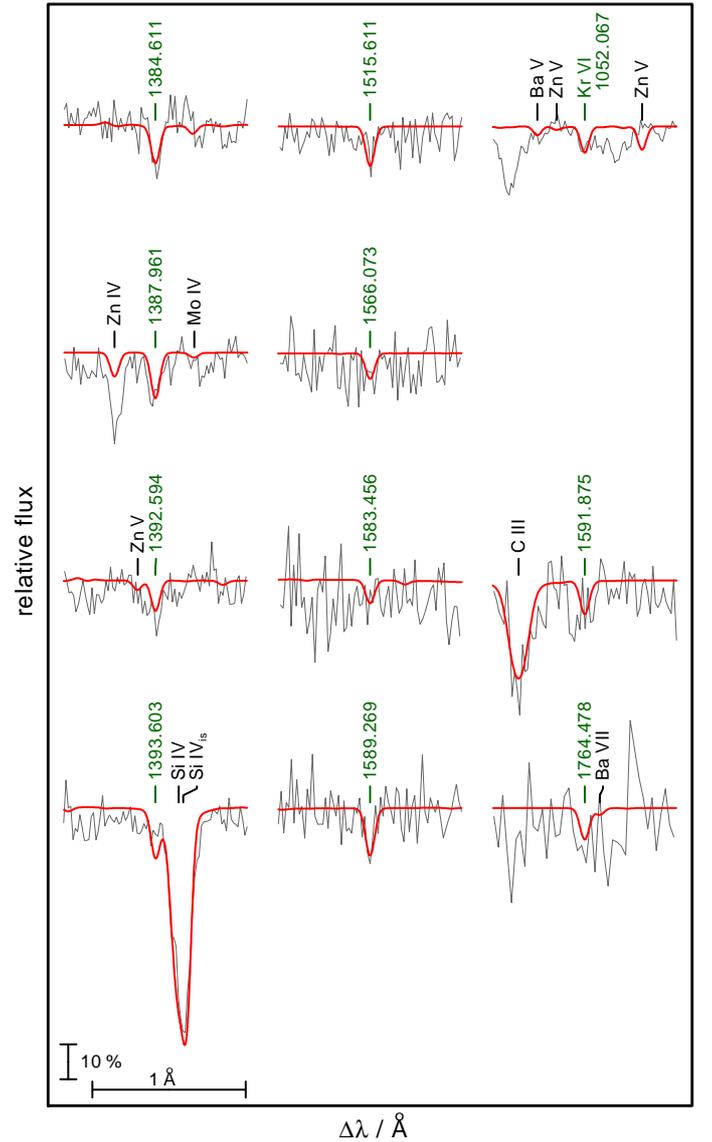}}
    \caption{Newly identified \ion{Kr}{v} lines and a \ion{Kr}{vi} (top, right) line in the FUSE and HST/STIS observations.
             The model is calculated with $\log \mathrm{Kr} = -3.3$.
            }
   \label{fig:krnew}
\end{figure}

Figure\,\ref{fig:ion} shows that \ion{Kr}{v-vii} are the dominant ions in the line-forming region.
We newly identified \ionw{Kr}{vi}{1052.067} ($\log gf = -0.55$) in the FUSE observation and for the first time
lines of \ion{Kr}{v}, namely 
$\lambda\lambda$
1384.611, 
1387.961, 
1392.594, 
1393.603,
1515.611,
1566.073,
1583.456,
1589.269,
1591.875,
1764.478 \AA\
($-$0.59, $-$0.65, $-$0.67, $-$0.17, 0.40, 0.30, 0.25, 0.68, 0.49, $-$0.16, respectively) in the
HST/STIS observation (Fig.\,\ref{fig:krnew}). These lines are in agreement with the observation,
while \ionww{Kr}{v}{1583.456, 1591.875, 1764.478} are uncertain. Many more weak \ion{Kr}{v-vii} lines are exhibited in the
synthetic spectrum  but they fade in the noise of the available observed UV and optical spectra.
The \ion{Kr}{v-vii} ionization equilibrium is well matched. Since ionization equilibria are sensitive indicators of
the effective temperature, our value of \Teffw{70\,000} (Sect.\,\ref{sect:tefflogg}) is corroborated.

With our new Kr oscillator strengths and also at the higher Kr abundance, a simultaneous fit of all 26 identified lines was achieved.
For example, \ion{Kr}{vi} $\lambda\lambda$ 944.046, 965.093, 1011.133, 1015.765, 1052.964\,\AA\ were much too weak before in our models
but now reproduce  the observation (Fig.\,\ref{fig:krwerner}).

In the optical wavelength range, \ionw{Kr}{v}{3579.739} ($\log gf = -1.15$, air wavelength 3578.717\,\AA) is the strongest line in our model 
but very weak (central depression of about 1\,\% of the local continuum flux) and not detectable in the available observation.

\paragraph{\ion{Ba}{vii}} lines were newly identified in the observed FUSE spectrum during the search for Kr lines.
These are one blend at \ionww{Ba}{vii}{924.892, 924.898} ($\log gf = -2.42$ and $-2.06$, respectively, Fig.\,\ref{fig:krfuse}),
and \ionw{Ba}{vii}{1143.317} ($-2.54$). \citet{rauchetal2014ba} previously discovered 
\ionw{Ba}{vii}{943.102} ($-1.77$) and
\ionw{Ba}{vii}{993.411} ($-1.57$).

\section{Mass, post-AGB age, and distance}
\label{sect:massdistance}

\begin{figure}
   \resizebox{\hsize}{!}{\includegraphics{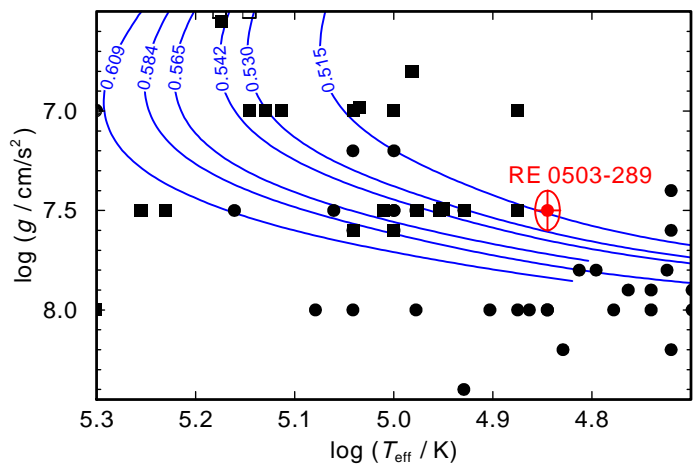}} 
    \caption{Location of \re in the
             $\log$ \Teff\,--\,\logg diagram (the ellipse indicates the error range)
             compared with evolutionary tracks for post-AGB stars that
             experienced a very late thermal pulse \citep{althausetal2009}.
             These are labeled with the respective stellar masses (in $M_\odot$). 
             Positions of hydrogen-deficient PG\,1159-type stars and DO-type WDs
             are indicated by squares and circles, respectively. 
             }
   \label{fig:evolre}
\end{figure}

\begin{figure}
   \resizebox{\hsize}{!}{\includegraphics{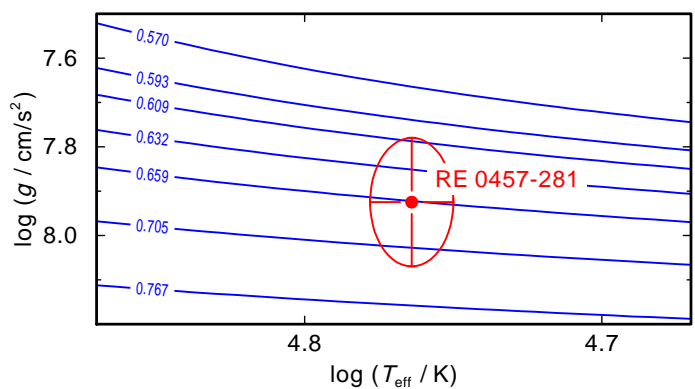}} 
    \caption{Location of \reb \citep[\Teffw{55\,875 - 60\,170}, \loggw{7.78 - 8.07},][the ellipse indicates the error range]{marshetal1997}
             in the $\log$ \Teff\,--\,\logg diagram 
             compared with evolutionary tracks for DA-type WDs \citep{renedoetal2010} 
             labeled with the respective stellar masses (in $M_\odot$). 
             }
   \label{fig:evolreb}
\end{figure}

We determined $M = 0.514^{+0.15}_{-0.05}\,\mathrm{M_\odot}$ from a comparison of the
evolutionary tracks of hydrogen-deficient post-AGB stars (Fig.\,\ref{fig:evolre}). From these
tracks, we calculated a post-AGB age of about $6.8 \pm \times 10^5$\,years.
For \reb, we measured $M = 0.660^{+0.65}_{-0.52}\,\mathrm{M_\odot}$ and a post-AGB age of
$1.14 \pm 0.06 \times 10^6$\,years by  comparing them with cooling sequences for old hydrogen-rich 
WDs (Fig.\,\ref{fig:evolreb}).

We used the flux calibration\footnote{\url{http://astro.uni-tuebingen.de/~rauch/SpectroscopicDistanceDetermination.gif}} of \citet{heberetal1984}
to calculate the distance
\begin{equation*}
d = 7.11 \times 10^4 \times \sqrt{H_\nu M \times 10^{0.4\, m_{\mathrm{V}_0}-\log g}}\,\mathrm{pc}\,,
\label{eq:distance}
\end{equation*}
\noindent
using
$m_\mathrm{V_0} = m_\mathrm{V} - R_\mathrm{v}E_\mathrm{B-V}$, and the Eddington flux 
$H_\nu = 1.018 \pm 0.002 \times 10^{-3}\, \mathrm{erg/cm^{2}/s/Hz}$ at $\lambda_\mathrm{eff} = 5454\,\mathrm{\AA}$
of our final model atmosphere.
We used 
$\ebv =0.015 \pm 0.002$, 
$M = 0.514^{+0.015}_{-0.005}$\,\Msol, and
$m_\mathrm{V} = 13.58 \pm 0.01$ \citep{faedietal2011}.
We derive
$d = 147^{+16}_{-18}$\,pc. 
The height below the Galactic plane\footnote{Galactic coordinates of \re for J2000: $l = 230\fdg 6717$, $b = -34\fdg 9355$} 
is $z = 84^{+9}_{-10}$\,pc. 
This distance is smaller than the value of \citet[][190\,pc]{vennesetal1998}.

\section{Interstellar line absorption}
\label{sect:ism}

\begin{figure}
   \resizebox{\hsize}{!}{\includegraphics{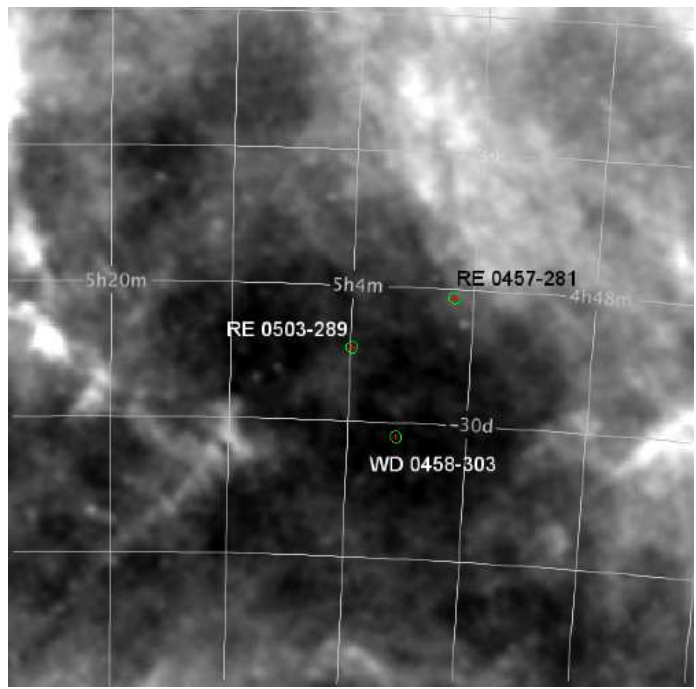}} 
    \caption{Locations (J2000) of \re, \reb, and \wda (marked by green encircled red + signs) in a $10\degr\,\times\,10\degr$ $100\,\mu$ dust map 
             from {\tt http://irsa.ipac.caltech.edu/applications/DUST}.
             }
   \label{fig:dust}
\end{figure}

To measure the interstellar reddening in the line of sight (LOS), we first normalized our synthetic spectrum (\Teffw{70\,000}, \loggw{7.5})
to match the measured $m_\mathrm{H} = 14.77$ \citep{cutrietal2003}. Then, interstellar reddening with $E_\mathrm{B-V} = 0.015 \pm 0.002$ had 
to be applied to reproduce the observed FUV continuum flux, using the reddening law of \citet{fitzpatrick1999} with the standard $R_\mathrm{v}=3.1$.
Our $E_\mathrm{B-V}$ value is in good agreement with measurements of 
\citet[][evaluating images from the Diffuse Infrared Background Experiment on board of the Cosmic Background Explorer satellite, COBE/DIRBE, 
                            and the Infrared Astronomy Satellite Sky Survey Atlas, IRAS/ISSA]{schlegeletal1998} and 
\citet[][based on Sloan Digital Sky Survey, SDSS, stellar spectra]{schlaflyfinkbeiner2011}.
They publish 
$E_\mathrm{B-V} = 0.0160$ and $E_\mathrm{B-V} = 0.0138$, respectively.
The errors can be estimated from the mean values of $E_\mathrm{B-V}$ within  a 5\degr\ circle around \re. These are
$E_\mathrm{B-V}^{5\degr} = 0.0155 \pm 0.0008$ and $0.0134 \pm 0.0006$, respectively. The dust distribution around \re and \reb is
illustrated by Fig.\,\ref{fig:dust}. While \re is apparently located in the middle of a voided area, \reb lies at the rim
of stronger emission, with a higher $E_\mathrm{B-V} = 0.0201$ 
\citep[$E_\mathrm{B-V}^{5\degr} = 0.0203 \pm 0.0004$,][]{schlaflyfinkbeiner2011}.

The DAO-type \wda 
\citep[\object{MCT\,0458$-$3020},
       $m_\mathrm{B} = 16.3$,
       \Teffw{91\,010 \pm 3156}, 
       \loggw{7.09 \pm 0.10},
       $M =  0.53 \pm 0.02\,M_\odot$,
       $d = 928\,\mathrm{pc}$,][]{demersetal1986,mccooksion1999,mccooksion1999cat,gianninasetal2010,gianninasetal2011}
also lies  close to \re (angular distance $1\degr 53$, Fig.\,\ref{fig:dust}) in an area with obviously less $100\,\mu$ emission and a lower
$E_\mathrm{B-V} = 0.0082$ 
\citep[$E_\mathrm{B-V}^{5\degr} = 0.0085 \pm 0.0005$,][]{schlaflyfinkbeiner2011}.

While \re and \reb were newly identified in the ROSAT/WFC (R\"ontgensatellit/Wide Field Camera) extreme-ultraviolet (EUV)
bright source catalogue \citep{poundsetal1993,poundsetal1993cat} and were later matched with their optical counterparts
\citep{masonetal1995,masonetal1996cat}, the much hotter \wda has no significant EUV flux. Therefore,
an investigation, based on UV spectroscopy, of the ISM line absorption in the LOS toward this much more distant star  is highly desirable.
So far, only Galaxy Evolution Explorer (GALEX\footnote{\url{http://www.galex.caltech.edu}}) near and far UV imaging is available in MAST. 
Exploiting the GALEX GR6 and GR7 data releases, GalexView\footnote{\url{http://galex.stsci.edu/GalexView}} provides
$m_\mathrm{FUV} = 14.35 \pm 0.01$, $m_\mathrm{NUV} = 15.08 \pm 0.01$ and $E_\mathrm{B-V} = 0.0096$ for \wda.

\begin{figure*}
   \resizebox{\hsize}{!}{\includegraphics{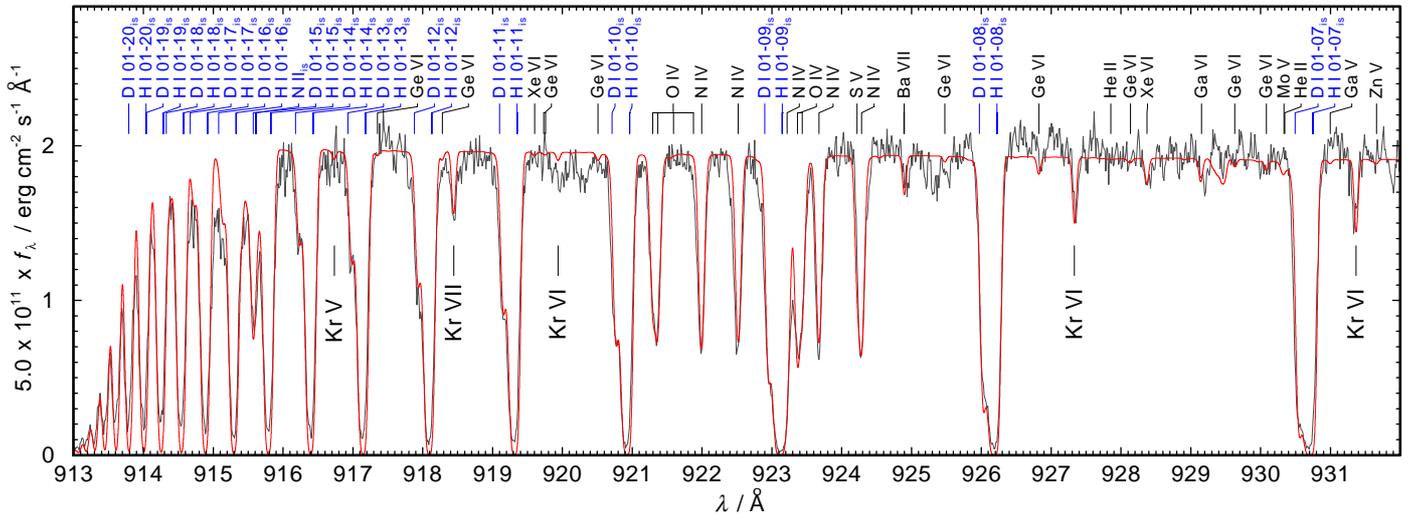}}
    \caption{Section of the FUSE observation (shifted to rest wavelengths) compared with our final synthetic spectrum (\Teffw{70\,000}, \loggw{7.5}).
             Prominent stellar and interstellar (IS, blue) lines are indicated.
            }
   \label{fig:krfuse}
\end{figure*}

Figure\,\ref{fig:krfuse} shows a section of the FUSE observation compared with our spectrum that was calculated with the new Kr model atom and 
$\log \mathrm{Kr} = -3.3$. \ionw{Kr}{vii}{918.444} and \ionww{Kr}{vi}{927.334, 931.368} are prominent in the observed spectrum and 
are well reproduced, while \ionw{Kr}{v}{916.734} and \ionw{Kr}{vi}{919.938}  are weak in our model and fade in the noise of the observation.

The program \emph{OWENS} was used to model the line absorption by the interstellar medium (ISM).
\emph{OWENS} is able to consider individual ISM clouds with different radial and turbulent velocities, temperatures, column
densities, and chemical compositions. Voigt profiles are fitted to the observation using a $\chi^2$ minimization.
More details are given by, for example, \citet{hebrard02} or \citet{hebrard03}.
To model the interstellar absorption of neutral hydrogen, we first considered  two clouds with column densities of
$N_\mathrm{\ion{H}{i}} =  1.2 \times 10^{18}\,\mathrm{cm^{-2}}$ and 
$N_\mathrm{\ion{H}{i}} =  9.3 \times 10^{15}\,\mathrm{cm^{-2}}$, and radial velocities of
$v^\mathrm{ISM}_\mathrm{rad} = +10.0$ and $-39.3\,\mathrm{km\,s^{-1}}$, respectively. 
These column densities are smaller than the expected value of $9.2 \pm 0.3 \times 10^{18}\,\mathrm{cm^{-2}}$
that was calculated from $N_\mathrm{H}\,/\,E_\mathrm{B-V} = 6.12 \pm 0.20 \times 10^{21}\,\mathrm{atoms\,cm^{-2}\,mag^{-1}}$
\citep[][with $N_\mathrm{H} = N_\ion{H}{i} + 2N(\mathrm{H_2})$]{gudennavaretal2012}.
In addition, we derive $v^\mathrm{RE\,0503-289}_\mathrm{rad} - v^\mathrm{ISM}_\mathrm{rad} = 14.5 \pm 4.2\,\mathrm{km\,s^{-1}}$ and 
$64.8 \pm 4.2\,\mathrm{km\,s^{-1}}$ for the two clouds. 

\citet{vennesetal1994} analyzed Extreme Ultraviolet Explorer (EUVE) photometry data and measured column densities of 
$\log (N_\mathrm{\ion{H}{i}}\,/\,\mathrm{cm^{-2}}), =  17.75 - 18.00$ 
and $17.80 - 17.90$ in the LOS toward \re and the nearby (spatially separated by $1\fdg 66$) \reb. 
Since \citet{hoareetal1993} and \citet{vallergaetal1993} determined
$\log (N_\mathrm{\ion{H}{i}}\,/\,\mathrm{cm^{-2}}) =  18.00 - 18.18$ for $\beta$ and $\epsilon$\,CMa  
(located in about the same direction, at angular distances of 21\degr\ and 31\degr, respectively,
but at larger distances of $d = 206\,\mathrm{pc}$ and $d = 188\,\mathrm{pc}$, respectively), \citet{vennesetal1994} suggested 
that the local cloud, agglomerated with a few parsecs from the Sun, is the main ISM structure along the LOS toward these stars. \citet{vennesetal1998} used ORFEUS/BEFS\footnote{Orbiting and Retrievable Far and Extreme Ultraviolet Spectrometer / Berkeley Extreme and Far-ultraviolet Spectrometer} 
observations and measured 
$v^\mathrm{RE\,0503-289}_\mathrm{rad} - v^\mathrm{ISM}_\mathrm{rad} = 48 \pm 21\,\mathrm{km\,s^{-1}}$, which is 
within error limits in agreement with the mean velocity of our two clouds.
\citet{dupuisetal1995} investigated interstellar column densities based on EUVE spectra.
For \reb, they found $d = 90\,\mathrm{pc}$ and $\log (N_\mathrm{\ion{H}{i}}\,/\,\mathrm{cm^{-2}}) =  18.04 - 18.12$. ) measured
$v^\mathrm{RE\,0457-281}_\mathrm{rad} = 80 \pm 12\,\mathrm{km\,s^{-1}}$ .
\citet{paulietal2006} investigated on the 3D kinematics of WDs from the SPY project and determined
$v^\mathrm{RE\,0457-281}_\mathrm{rad} = 49.2 \pm 11.5\,\mathrm{km\,s^{-1}}$ and a distance of $d = 115.9 \pm 14\,\mathrm{pc}$.

\begin{figure*}
   \resizebox{\hsize}{!}{\includegraphics{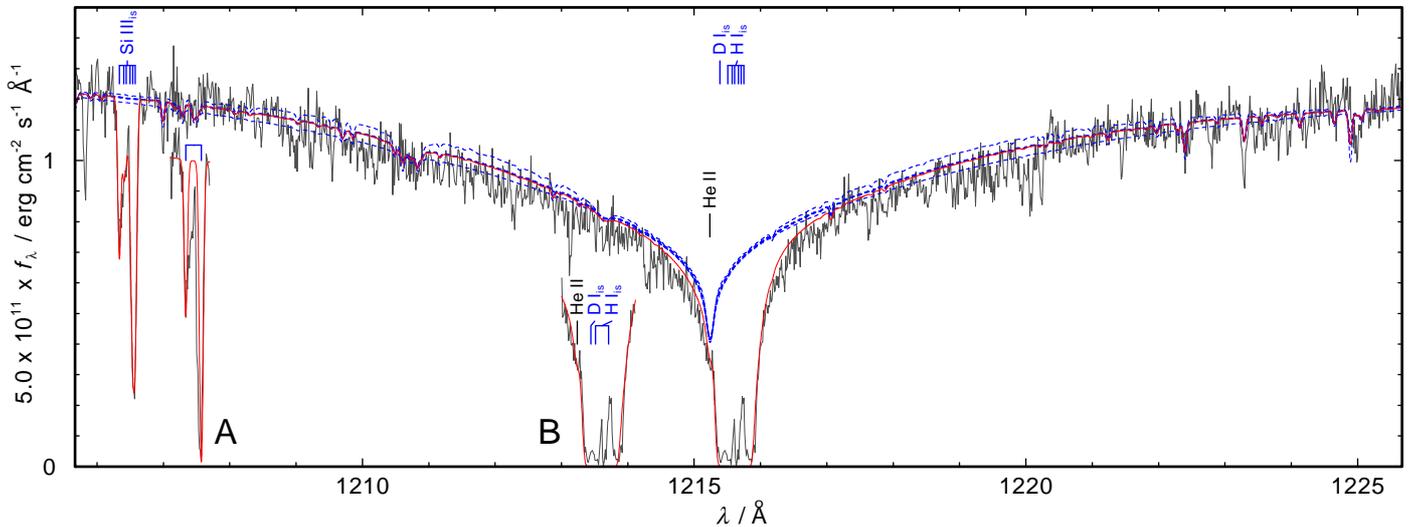}}
    \caption{Section of the STIS observation around L\,$\alpha$, compared with our final synthetic spectrum (thick, red, \Teffw{70\,000}, \loggw{7.5}).
             The dashed, blue lines are the pure photospheric model-atmosphere spectra (\Teffw{70\,000}, \loggw{7.0, 7.5, 8.0}), 
             i.e., no interstellar line absorption is applied. See text for the description of inserts A and B.
            }
   \label{fig:stisism}
\end{figure*}

The interstellar \ionw{N}{ii}{915.6} also exhibits  a double feature (Fig.\,\ref{fig:krfuse}).
We considered this line with column densities of 
$N_\mathrm{\ion{N}{ii}} = 7.5 \times 10^{13}\mathrm{cm^{-2}}$ at $+11.0\,\mathrm{km\,s^{-1}}$ and
$N_\mathrm{\ion{N}{ii}} = 2.5 \times 10^{13}\mathrm{cm^{-2}}$ at $-39.5\,\mathrm{km\,s^{-1}}$. 
The HST/STIS observation is used to verify our solution with two ISM clouds (Fig.\,\ref{fig:stisism}). 
While \ionw{Si}{iii}{1206.5} is not sufficiently well reproduced with
$N_\mathrm{\ion{S}{iii}} = 1.8 \times 10^{13}\mathrm{cm^{-2}}$ at $+18.0\,\mathrm{km\,s^{-1}}$ and
$N_\mathrm{\ion{S}{iii}} = 1.9 \times 10^{12}\mathrm{cm^{-2}}$ at $-40.5\,\mathrm{km\,s^{-1}}$ (insert A),
a multi-cloud solution for the ISM absorption could improve the agreement (main spectra). 
The assumed clouds' column densities and radial velocities are summarized in Table\,\ref{tab:is}.

\begin{table*}\centering
\caption{Ionic column densities (in $\mathrm{cm^{-2}}$) and radial velocities (in $\mathrm{km\,s^{-1}}$) in interstellar clouds in the line of sight toward \re.}         
\label{tab:is}
\setlength{\tabcolsep}{.16em}
\begin{tabular}{cr@{.}lccr@{.}lccr@{.}lccr@{.}lccr@{.}lccr@{.}lccr@{.}l}
\hline
\hline
\noalign{\smallskip}
\multicolumn{3}{c}{\ionw{C}{ii}{1036.3}} && 
\multicolumn{3}{c}{\ionw{C}{iii}{977.0}} && 
\multicolumn{3}{c}{\ionw{N}{ii}{1084.0}} && 
\multicolumn{3}{c}{\ionw{O}{i}{988.6}} && 
\multicolumn{3}{c}{\ionw{O}{vi}{1031.9}} && 
\multicolumn{3}{c}{\ionw{Si}{ii}{1260.4}} && 
\multicolumn{3}{c}{\ionw{Si}{iii}{1206.5}} \\
\multicolumn{3}{c}{\ionw{C}{ii}{1334.5}} && 
\multicolumn{3}{c}{} && 
\multicolumn{3}{c}{} && 
\multicolumn{3}{c}{\ionw{O}{i}{988.7}} && 
\multicolumn{3}{c}{\ionw{O}{vi}{1037.6}} && 
\multicolumn{3}{c}{\ionw{Si}{ii}{1526.7}} && 
\multicolumn{3}{c}{} \\
\multicolumn{3}{c}{} && 
\multicolumn{3}{c}{} && 
\multicolumn{3}{c}{} && 
\multicolumn{3}{c}{\ionw{O}{i}{988.8}} && 
\multicolumn{3}{c}{} && 
\multicolumn{3}{c}{} && 
\multicolumn{3}{c}{} \\
\cline{1-3}
\cline{5-7}
\cline{9-11}
\cline{13-15}
\cline{17-19}
\cline{21-23}
\cline{25-27}
\noalign{\smallskip}
$N$ & \multicolumn{2}{c}{$v_\mathrm{rad}$} && 
$N$ & \multicolumn{2}{c}{$v_\mathrm{rad}$} && 
$N$ & \multicolumn{2}{c}{$v_\mathrm{rad}$} &&
$N$ & \multicolumn{2}{c}{$v_\mathrm{rad}$} &&
$N$ & \multicolumn{2}{c}{$v_\mathrm{rad}$} &&
$N$ & \multicolumn{2}{c}{$v_\mathrm{rad}$} && 
$N$ & \multicolumn{2}{c}{$v_\mathrm{rad}$} \\
\hline
\noalign{\smallskip}
$7.0 \times 10^{13}$ & $+18$&$0$ &&  $8.0 \times 10^{12}$ & $+17$&$0$ &&  $3.0 \times 10^{13}$ & $+17$&$0$ &&  $8.0 \times 10^{13}$ & $+19$&$0$ &&  $1.0 \times 10^{13}$ & $+17$&$0$ &&  $5.0 \times 10^{12}$ & $+17$&$0$ &&  $4.0 \times 10^{12}$ & $+19$&$0$ \\
$7.0 \times 10^{13}$ & $ +8$&$0$ &&  $8.0 \times 10^{12}$ & $ +7$&$0$ &&  $3.0 \times 10^{13}$ & $ +7$&$0$ &&  $5.0 \times 10^{13}$ & $ +9$&$0$ &&  $1.0 \times 10^{13}$ & $ +7$&$0$ &&  $5.0 \times 10^{12}$ & $ +9$&$0$ &&  $4.0 \times 10^{12}$ & $ +9$&$0$ \\
$1.0 \times 10^{13}$ & $ -0$&$5$ &&  $8.0 \times 10^{12}$ & $ -0$&$5$ &&  $1.0 \times 10^{13}$ & $ -0$&$5$ &&  $5.0 \times 10^{13}$ & $ -0$&$5$ &&  $1.0 \times 10^{13}$ & $ -0$&$5$ &&  $1.5 \times 10^{12}$ & $ -0$&$5$ &&  $6.0 \times 10^{11}$ & $ -0$&$5$ \\
$9.0 \times 10^{12}$ & $-14$&$5$ &&  $6.0 \times 10^{12}$ & $-14$&$5$ &&  $7.0 \times 10^{12}$ & $-14$&$5$ &&  $4.0 \times 10^{13}$ & $-14$&$5$ &&  $9.0 \times 10^{12}$ & $-14$&$5$ &&  $5.0 \times 10^{11}$ & $-14$&$5$ &&  $4.5 \times 10^{11}$ & $-14$&$5$ \\
$9.0 \times 10^{12}$ & $-24$&$5$ &&  $4.0 \times 10^{12}$ & $-24$&$5$ &&  $4.0 \times 10^{12}$ & $-24$&$5$ &&  $5.0 \times 10^{13}$ & $-24$&$5$ &&  $9.0 \times 10^{12}$ & $-24$&$5$ &&  $2.5 \times 10^{11}$ & $-24$&$5$ &&  $8.0 \times 10^{11}$ & $-24$&$5$ \\
$2,2 \times 10^{13}$ & $-40$&$5$ &&  $6.0 \times 10^{12}$ & $-40$&$5$ &&  $1.5 \times 10^{13}$ & $-40$&$5$ &&  $4.0 \times 10^{13}$ & $-40$&$5$ &&  $1.7 \times 10^{13}$ & $-40$&$5$ &&  $1.5 \times 10^{12}$ & $-40$&$5$ &&  $1.9 \times 10^{12}$ & $-40$&$5$ \\
\hline
\end{tabular}
\end{table*}

While the outer line wings of \ion{H}{i} L\,$\alpha$ are dominated by stellar \ion{He}{ii} absorption,
the inner line wings are well matched at a total $N_\mathrm{\ion{H}{i}} =  1.5 \times 10^{18}\,\mathrm{cm^{-2}}$.
To reproduce the blue side of the observed absorption core, a significant \ion{D}{i} column density
of $N_\mathrm{\ion{D}{i}} = 1.2 \times 10^{14}\,\mathrm{cm^{-2}}$ is necessary. This value is uncertain because 
this line is saturated and blended by stellar \ion{He}{ii} and interstellar \ion{H}{i}. 
Therefore, we consider \ion{D}{i} with a single ISM cloud at $12.1\,\mathrm{km\,s^{-1}}$.
The region around L\,$\alpha$ is well reproduced with either a two-cloud (insert ``B') or a multi-cloud 
solution for \ion{H}{i}. 

\begin{figure*}
   \resizebox{\hsize}{!}{\includegraphics{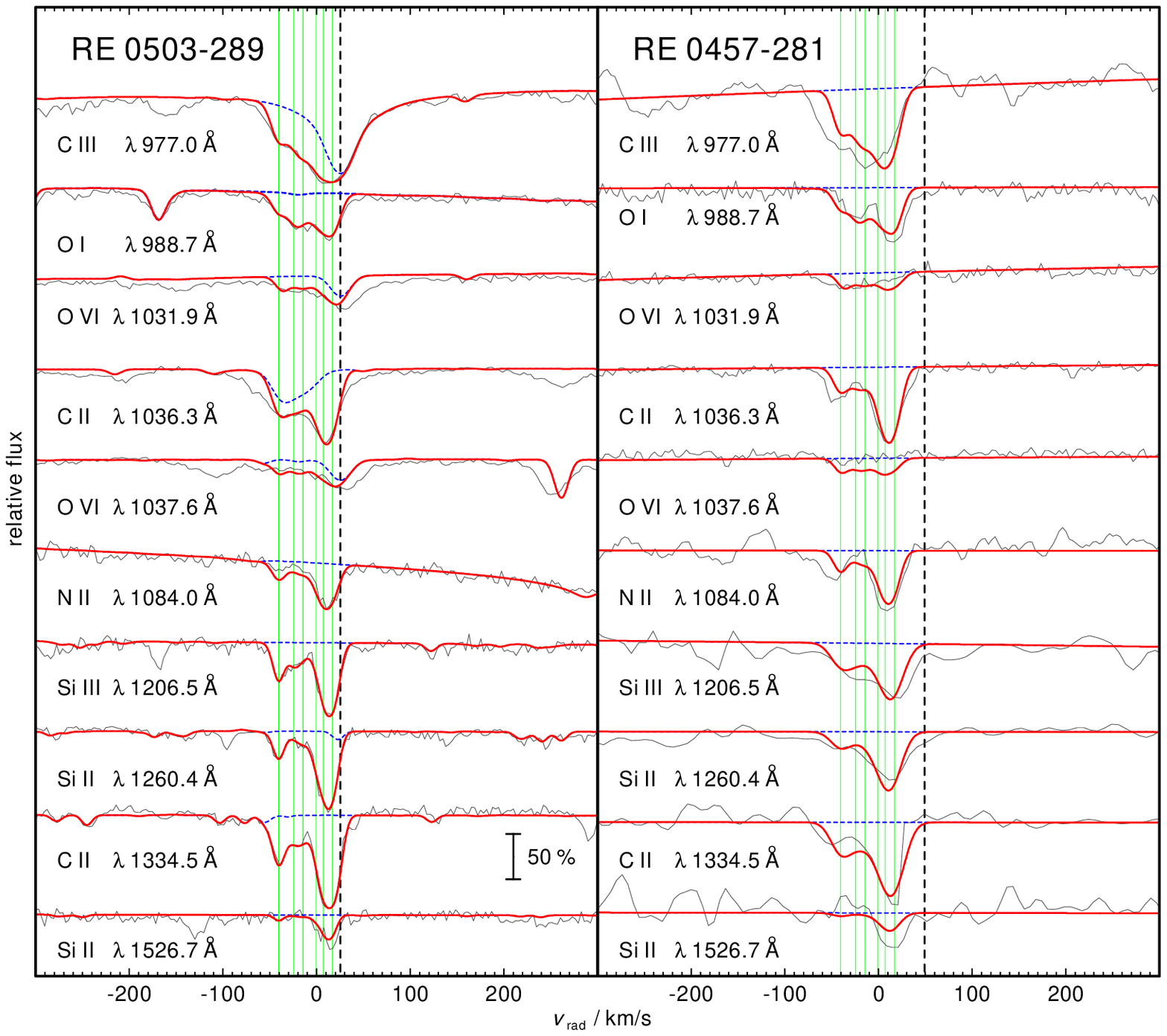}}
    \caption{Sections of the FUSE, STIS, and IUE observations around interstellar lines
             compared with our synthetic spectra for \re (left) and \reb (right).
             The pure stellar model spectra are shown with dashed, blue lines. 
             The vertical lines indicate the assumed clouds' velocities given in Table\,\ref{tab:is}.
             The dashed, vertical lines in each panel show the radial velocities of the two stars.
            }
   \label{fig:ism}
\end{figure*}

\begin{figure}
   \resizebox{\hsize}{!}{\includegraphics{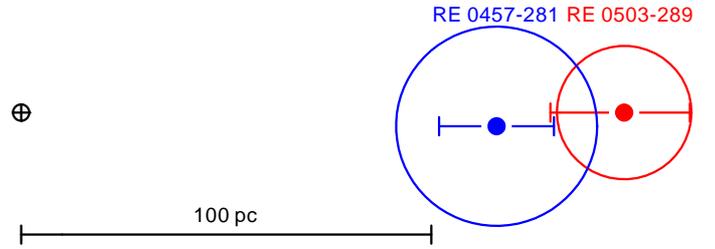}}
    \caption{Constellation of Earth, \reb, and \re.
             The circles indicate the estimated maximum distance reached by
             stellar material that was ejected from the stars on the AGB.
            }
   \label{fig:re0457_0503}
\end{figure}

Figure\,\ref{fig:ism} shows that a multi-cloud solution can explain the strong and weak ISM absorption lines in the
observations of \re and \reb. To simulate the stellar flux of \reb,
we used a synthetic spectrum (pure hydrogen, \Teffw{58\,000}, \loggw{7.9}) which was provided by the 
German Astrophysical Virtual Observatory (GAVO\footnote{\url{http://www.g-vo.org}}) Theoretical Stellar Spectra Access service 
(TheoSSA\footnote{\url{http://dc.g-vo.org/theossa}}).
The ISM line absorption was modeled with the same parameters that were used for \re. 
We find a good agreement for \reb and \re, although they have different distances of
$116 \pm 14\,\mathrm{pc}$ and 
$150^{+17}_{-18}\,\mathrm{pc}$, respectively, which are in rough agreement within their error limits (Fig.\,\ref{fig:re0457_0503}).
This issue will be clarified by the results of GAIA\footnote{\url{http://sci.esa.int/gaia}} in the near future.

This corroborates the suggestion of \citet{vennesetal1994} that the major contribution to the ISM absorption stems 
from the so-called local fluff in which our Sun is located. Nearly all of the gas along the LOS toward \re and
\reb, and at distances well beyond these, is very hot and highly ionized. The gas that we see occupies only a small fraction 
of the total distance to the stars. 

The main interstellar gas component toward \re is located
near $v_{\rm helio}=+15\,\mathrm{km\,s^{-1}}$, as evident from the \ion{H}{i}
21\,cm emission spectrum in this direction from the Leiden-Argentina-Bonn
(LAB) survey \citep{kalberlaetal2005}.
Also the STIS spectrum of \re indicates strong interstellar
absorption at $v_\mathrm{helio}\approx +15\,\mathrm{km\,s^{-1}}$ in the strong
resonance lines of \ionw{C}{ii}{1334.5}, \ionw{Si}{iii}{1206.5},
\ionw{Si}{ii}{1260.4}, and others. As well as this main absorption
component, there is weaker interstellar/circumstellar
absorption extending bluewards until $-60\,\mathrm{km\,s^{-1}}$,
including another distinct (weak) absorption component near $v_\mathrm{helio}=-40\,\mathrm{km\,s^{-1}}$
(Fig.\,\ref{fig:ism}). 
The strong \ionw{C}{ii}{1334.5} and \ionw{Si}{iii}{1206.5} lines show additional
weak absorption between $-30\,\mathrm{km\,s^{-1}}$ and zero velocities, but without a clear
component structure.

The \ion{H}{i} Lyman series absorption can also  be best fit  with two
neutral gas components at $+15$ and $-40\,\mathrm{km\,s^{-1}}$, with a total column density
of $N_\mathrm{\ion{H}{i}} =  1.5 \times 10^{18}\,\mathrm{cm^{-2}}$.

In the metal ions, the satellite component at $-40\,\mathrm{km\,s^{-1}}$ is relatively
narrow with a $b$-value of just $2.5\,\mathrm{km\,s^{-1}}$. The simultaneous presence of weak
\ion{C}{ii}, \ion{Si}{ii}, \ion{Si}{iii}, and \ion{H}{i} absorption
(and possibly \ion{O}{vi}; Fig.\,\ref{fig:ism}), together with the narrow line shape,
indicates a relatively compact, low-column density gas structure with
multiphase gas that causes the absorption at negative velocities.

To estimate the contribution of the circumstellar material to the ISM absorption in
the LOS toward \re and \reb, we estimated the densities of
planetary nebulae (PNe) that were ejected at the end of the AGB phases of both
stars. We assumed expansion velocities of 
$v_\mathrm{exp} = 20\,\mathrm{km\,s^{-1}}$. Because of the very long post-AGB times,
these PNe have swept up all stellar material ejected in the slow ($\approx 10\,\mathrm{km\,s^{-1}}$) AGB-wind
phase before, and their so-called radii indicate the maximum distance from the star that ejected
material would have reached.
 
Table\,\ref{tab:pn} summarizes radii, volumes, masses, and densities of 
the expected PNe. The estimated column densities are orders of magnitude lower than those
that are necessary to reproduce the observation, even if \re lies beyond the 
circumstellar material of \reb. However, the ejected PN material may have compressed accelerated
ambient interstellar gas, so that the $-40\,\mathrm{km\,s^{-1}}$ component toward \re
may be a result of the interaction between circumstellar and interstellar material
at the interface between both components.

\begin{table}\centering 
  \caption{Parameters to estimate the circumstellar column densities around \re and \reb
           owing to their AGB mass loss.}
\label{tab:pn}
\setlength{\tabcolsep}{.4em}
\begin{tabular}{r@{\,/\,}lr@{.}lr@{.}l}
\hline
\hline
\noalign{\smallskip}                                                                                          
\multicolumn{2}{c}{} & \multicolumn{2}{c}{\re} & \multicolumn{2}{c}{\reb} \\
\hline
\noalign{\smallskip}                                                                                          
Post-AGB age & $10^6$\,a                           &  0&80\tablefootmark{a}                    &   1&2\tablefootmark{b}                      \\
\noalign{\smallskip}                                                                                          
$R_\mathrm{max}^\mathrm{PN}$ & cm                   &  5&$0 \times 10^{19}$\ \ \tablefootmark{c} & $7$&$5 \times 10^{19}$\ \ \tablefootmark{c} \\
\noalign{\smallskip}                                                                                          
$R_\mathrm{max}^\mathrm{PN}$ & pc                   & 16&4                                      &  24&5                                       \\
\noalign{\smallskip}                                                                                          
$V_\mathrm{max}^\mathrm{PN}$ & cm$^3$               &  5&$3 \times 10^{59}$                      & $1$&$8 \times 10^{60}$                       \\
\noalign{\smallskip}                                                                                          
$M_\mathrm{ZAMS\tablefootmark{d}}$ & $M_\odot$     &  1&0\tablefootmark{a}                      &   2&5\tablefootmark{b}                      \\
\noalign{\smallskip}                                                                                          
$M_\mathrm{final}$ & $M_\odot$                     &  0&514\tablefootmark{e}                    &   0&660\tablefootmark{f}                    \\
\noalign{\smallskip}                                                                                          
AGB mass loss & $M_\odot$                          &  0&486                                    &   1&840                                     \\
\noalign{\smallskip}                                                                                          
AGB mass loss & $M_\ion{H}{i}$                     &  5&$8 \times 10^{56}$                     & $2$&$2 \times 10^{57}$                       \\
\noalign{\smallskip}                                                                                          
Column density\tablefootmark{g} &  cm$^{-2}$          &  5&$4 \times 10^{15}$                     & $9$&$2 \times 10^{15}$                       \\
\hline
\end{tabular}
\tablefoot{
\tablefoottext{a}{\citet{althausetal2009}}
\tablefoottext{b}{\citet{renedoetal2010}}
\tablefoottext{c}{$v_\mathrm{exp} = 20\,\mathrm{km\,s^{-1}}$ assumed}
\tablefoottext{d}{zero-age main sequence}
\tablefoottext{e}{\citet{werneretal2015}}
\tablefoottext{f}{this paper}
\tablefoottext{g}{circumstellar, assumed location of swept up material within 0.9 - 1.0 $R_\mathrm{max}^\mathrm{PN}$}
}
\end{table}

\section{Results and conclusions}
\label{sect:results}

We reanalyzed the effective temperature and surface gravity and determined
\Teffw{70\,000 \pm 2000} and 
\loggw{7.5 \pm 0.1}.
This verifies the results of \citet{dreizlerwerner1996} within improved, rather small, error limits.

For precise NLTE spectral analyses, reliable transition probabilities are required, not only for lines that are
identified in the observation, but also for the complete model atoms that are considered in the model-atmosphere calculations.
Therefore, our new computation of a complete set of transition probabilities for \ion{Kr}{iv-vii} transition probabilities
was a prerequisite for an improved NLTE spectral analysis. The new data enabled us to construct a more realistic Kr model
atom. We improved the previous determination of the Kr abundance in \re \citep{werneretal2012}, taking  these oscillator strengths into consideration.

\begin{figure}
   \resizebox{\hsize}{!}{\includegraphics{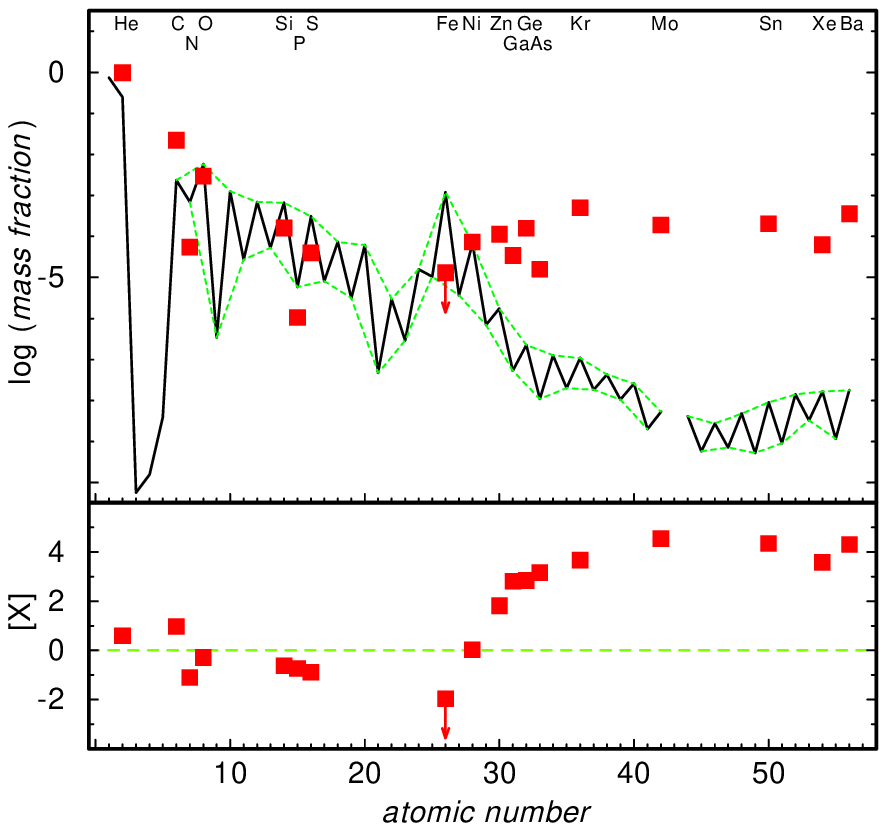}}
    \caption{Solar abundances \citep[thick line; the dashed lines
             connect the elements with even and  odd atomic numbers]{asplundetal2009,scottetal2015a,scottetal2015b,grevesseetal2015}
             compared with the determined photospheric abundances of 
             \re \citep[red squares,][and this work]{dreizlerwerner1996,rauchetal2014zn, rauchetal2015ga, rauchetal2012ge, rauchetal2016mo, rauchetal2014ba}.
             The uncertainties of the WD abundances are about 0.2\,dex, in general. For Fe, the arrow indicates the upper limit.
             Top panel: Abundances given as logarithmic mass fractions.
             Bottom panel: Abundance ratios to respective solar values, 
                           [X] denote log (fraction\,/\,solar fraction) of species X.
                           The dashed, green line indicates solar abundances.
            }
   \label{fig:X}
\end{figure}

In addition to the already known 15 \ion{Kr}{vi-vii} lines in the observed high-resolution UV spectrum of \re \citep{werneretal2012}, 
we identified one \ion{Kr}{vi} line and, for the first time in this star, ten lines of \ion{Kr}{v}. 
Our synthetic line profiles reproduce well the observation at a photospheric Kr abundance of
$2.5 \times 10^{-4} - 1.0 \times 10^{-3}$ ($\log \mathrm{Kr} = -3.3 \pm 0.3$). 
This is $2300 - 9200$ times the solar abundance \citep{grevesseetal2015}.
This highly supersolar Kr abundance goes along with the high abundances of other trans-iron elements in \re (Fig.\,\ref{fig:X}).
The \ion{Kr}{v-vii} ionization equilibrium is well reproduced (Figs.\,\ref{fig:krwerner}, \ref{fig:krnew}).

\begin{figure}
   \resizebox{\hsize}{!}{\includegraphics{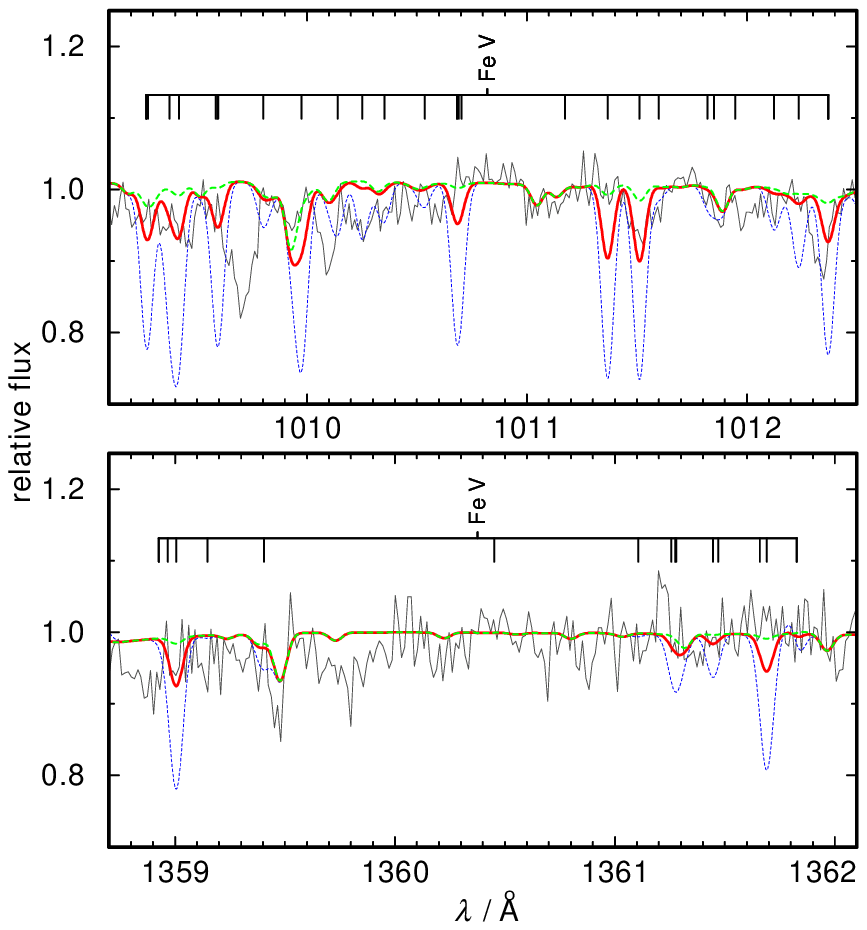}}
    \caption{Sections of the FUSE (top panel) and HST/STIS (bottom) spectra compared with synthetic spectra
             (\Teffw{70\,000}, \loggw{7.5}) that were calculated with Fe abundances of
             $1.3 \times 10^{-5}$ (dashed, green),
             $1.3 \times 10^{-4}$ (full, red), and
             $1.3 \times 10^{-3}$ (dashed, blue) (about 0.01, 0.1, 1 times solar, respectively).
             \ion{Fe}{v} lines are indicated.
            }
   \label{fig:fe}
\end{figure}

Iron is the only element in Fig.\,\ref{fig:X} with an upper abundance limit 
\citep[Fe/He $< 10^{-6}$ by number, about a hundredth of the solar abundance,][]{scottetal2015b}. \citet{barstowetal2000} 
determined this value by a co-addition (in the velocity space) of the nine \ion{Fe}{v} lines that were 
predicted to be strongest in the HST/GHRS spectrum. In Fig\,\ref{fig:fe}, we compare theoretical line
profiles of the most prominent \ion{Fe}{v} lines in the FUSE and HST/STIS wavelength range to the
observation. The upper limit for the Fe abundance of 0.01 times solar is verified. Therefore, the reason why the Ni/Fe abundance ratio is much higher compared to other post-AGB stars
remains unexplained \citep[see][]{barstowetal2000}.

\begin{acknowledgements}
TR and DH are supported by the German Aerospace Center (DLR, grants 05\,OR\,1507 and 50\,OR\,1501, respectively).
The GAVO project had been supported by the Federal Ministry of Education and
Research (BMBF) 
at T\"ubingen (05\,AC\,6\,VTB, 05\,AC\,11\,VTB) and is funded
at Heidelberg (05\,AC\,11\,VH3).
Financial support from the Belgian FRS-FNRS is also acknowledged. 
PQ is research director of this organization.
Some of the data presented in this paper were obtained from the
Mikulski Archive for Space Telescopes (MAST). STScI is operated by the
Association of Universities for Research in Astronomy, Inc., under NASA
contract NAS5-26555. Support for MAST for non-HST data is provided by
the NASA Office of Space Science via grant NNX09AF08G and by other
grants and contracts. 
We thank 
Ralf Napiwotzki for putting the reduced ESO/VLT spectra at our disposal,
Monica Raineri who sent us the electronic versions of the
\ion{Kr}{iv} \citep{brediceetal2000} and
\ion{Kr}{v} \citep{rainerietal2012} $\log gf$ data, and
Liang Liang who provided the
\ion{Kr}{vii} data \citep{liangetal2013}.
This work used the profile-fitting procedure, OWENS, that was developed by M\@. Lemoine and the FUSE French Team.
This research has made use of 
NASA's Astrophysics Data System and
the SIMBAD database, operated at CDS, Strasbourg, France.
The TheoSSA service (\url{http://dc.g-vo.org/theossa}) used to retrieve theoretical spectra for this paper 
and the TOSS service (\url{http://dc.g-vo.org/TOSS}) that provides weighted oscillator strengths and transition probabilities
were constructed as part of the activities of the German Astrophysical Virtual Observatory.
\end{acknowledgements}

\bibliographystyle{aa}
\bibliography{28131}

\end{document}